\documentclass[final]{siamltex}
\usepackage{amsfonts}
\usepackage{amstex}
\usepackage{latexsym}
\usepackage{subeqn}
\usepackage{epsfig}

\newcommand{\tr}{{\rm tr}}
\newtheorem{remark}[theorem]{Remark}

\title{Approximation of the determinant of large sparse symmetric
positive definite matrices}

\author{Arnold Reusken\thanks{Institut f\"{u}r Geometrie und Praktische 
Mathematik, RWTH Aachen, Templergraben 55, D-52056 Aachen, Germany.}}

\begin{document}

\maketitle

\begin{abstract}
This paper is concerned with the  problem of approximating $\det(A)^{1/n}$ for a
 large sparse symmetric
positive definite matrix $A$ of order $n$. It is shown that an efficient
solution of this problem is obtained  by using a sparse approximate inverse
of $A$. The method is explained and theoretical properties are discussed.
A posteriori error estimation techniques are presented. Furthermore, results of
numerical experiments are given which illustrate the performance of this
new method.  
\end{abstract}



\pagestyle{myheadings}
\thispagestyle{plain}
\markboth{A. REUSKEN}{APPROXIMATION OF DETERMINANTS}

\section{Introduction}
Throughout this paper, $A$ denotes a real symmetric positive definite matrix
of order $n$ with eigenvalues
\[
  0< \lambda_1 \leq \lambda_2 \leq \ldots \leq \lambda_n.
\]
In a number of applications, for example in lattice Quantum Chromodynamics
\cite{Montvay}, certain functions of the determinant of $A$, such as
$\det(A)^{1/n}$ or $\ln(\det(A))$ are of interest. It is well-known (cf. also
\S\ref{Preliminaries}) that for large $n$ the function 
$A \rightarrow \det(A)$ has poor scaling properties and can be very
ill-conditioned for certain matrices $A$. In this paper we consider
the function
\begin{equation}
d:~A \rightarrow \det(A)^{\frac{1}{n}} ~.
\label{dfunction}
\end{equation}
A few basic properties of this function are discussed in \S\ref{Preliminaries}.
In this paper we present a new method for approximating $d(A)$ for large
sparse matrices $A$. The method is based on replacing $A$ by a matrix which
is in a certain sense close  to $A^{-1}$ and for which the
determinant can be computed with low computational costs. 
One popular method for approximating $A$  is
based on the construction  of an  incomplete Cholesky factorization.
This incomplete factorization is often used as a preconditioner
when solving linear systems with matrix $A$. In this paper
we use another preconditioning technique, namely that of sparse approximate
inverses (cf. \cite{Axelsson,Grote,Kaporin,Kol2}). In Remark~\ref{properties} we comment on the advantages
of the use of  sparse approximate inverse preconditoning for approximating
$d(A)$. Let $A=LL^T$ be the Cholesky decomposition of $A$. Then
using techniques known from the literature a sparse
approximate inverse $G_E$ of $L$, i.e. a lower triangular
 matrix $G_E$ which has a prescribed 
sparsity structure $E$ and which is an approximation of $L^{-1}$, can be constructed.
We then use $\det(G_E)^{-2/n}= \prod_{i=1}^n (G_E)_{ii}^{-2/n}$ as an approximation
 for $d(A)$. In \S\ref{SPAI} we explain the construction of $G_E$ and discuss
 theoretical properties of this sparse approximate inverse. For example, 
 such a sparse approximate inverse can be shown to exist for any symmetric
 positive definite $A$ and has an interesting optimality property
 related to $d(A)$.
 As a direct consequence of this optimality property one obtains that $d(A) \leq
 \det(G_E)^{-2/n}$ holds and that the approximation of $d(A)$ by $\det(G_E)^{-2/n}$
 becomes better if  a larger sparsity pattern $E$ is used. \\
 In \S\ref{ErrorEstimation} we consider the topic of error estimation.
 In the paper \cite{Bai} bounds for the determinant of symmetric positive definite
 matrices are derived. These bounds, in which the Frobenius norm
 and an estimate of the extreme eigenvalues
 of the matrix involved are used, often yield rather poor
 estimates of the determinant (cf. experiments in \cite{Bai}).
 In \S\ref{method1} we apply this technique to the preconditioned matrix
 $G_EAG_E^T$ and thus obtain reliable but rather pessimistic
 error bounds.  It turns out that this error estimation technique
 is rather
 costly. In \S\ref{method2} we introduce a
 simple and cheap Monte Carlo technique for error estimation.
  In \S\ref{Experiments} we apply the new method to a few examples of
 large
 sparse  symmetric positive definite matrices. 
 \section{Preliminaries}
 \label{Preliminaries}
 In this section we discuss a few elementary properties of the function $d$.
 We give a comparision between the conditioning of the function $d$ and
 of the fuction $A\rightarrow d(A)^n=\det(A)$. We use the notation
 $\| \cdot \|_2$ for the Euclidean norm and $\kappa(A)=\lambda_n/ \lambda_1$
 denotes the spectral condition number of $A$. The trace of the matrix $A$ is
 denoted
 by $\tr(A)$.
 \begin{lemma}
 \label{basics}
 Let $A$ and $\delta A$ be symmetric positive definite matrices of order $n$. The
 following inequalities hold:
 \begin{subequations}\label{basicABC}
 \begin{equation}\label{basicA}
  \lambda_1 \leq d(A) \leq \lambda_n~,
 \end{equation}
 \begin{equation}\label{basicB}
 d(A) \leq \frac{1}{n} \tr(A)~,
 \end{equation}
 \begin{equation}\label{basicC}
 0 < \frac{d(A+\delta A)-d(A)}{d(A)} \leq \kappa(A) \frac{\|\delta A\|_2}{\|A\|_2}~.
 \end{equation}
\end{subequations}
 \end{lemma} 
 
 \begin{proof}
 The result in (\ref{basicA}) follows from 
 \[
    \lambda_1 \leq (\prod_{i=1}^n \lambda_i)^{\frac{1}{n}} \leq \lambda_n~.
 \]
 The result in (\ref{basicB}) follows from the inequality between the
 geometric and arithmetic mean:
 \[
  d(A) = (\prod_{i=1}^n \lambda_i)^{\frac{1}{n}} \leq \frac{1}{n} \sum_{i=1}^n
   \lambda_i =\frac{1}{n} \tr(A) ~.
 \]
 From the Courant-Fischer characterization of eigenvalues it follows that
 \[
  \lambda_i(A+\delta A) \geq \lambda_i(A)+\lambda_1(\delta A) > \lambda_i(A)
 \]
 for all $i$. Hence $d(A+\delta A) > d(A) $ holds. Now note that
 \begin{eqnarray*}
  \frac{d(A+\delta A)-d(A)}{d(A)} & = &
    \left(\det(I+A^{-1}\delta A)\right)^{\frac{1}{n}}-1 \\ 
    & = & \left( \prod_{i=1}^n (1+ \lambda_i(A^{-1} \delta A))
    \right)^{\frac{1}{n}} -1 \\
    & \leq & \left( \prod_{i=1}^n (1+ \|A^{-1}\|_2 \|\delta A\|_2)
    \right)^{\frac{1}{n}} -1 \\
    &=& \|A^{-1}\|_2 \|\delta A\|_2 ~=~ \kappa(A) \frac{\|\delta A\|_2}{\|A\|_2}
    ~.
  \end{eqnarray*}
 Thus the result in (\ref{basicC}) is proved. 
 \qquad\end{proof}
 
 The result in (\ref{basicC}) shows that the function $d(A)$ is well-conditioned
 for matrices $A$ which have a not too
 large condition number $\kappa(A)$.
 
 We now briefly discuss the difference in conditioning between the functions
 $A \rightarrow d(A)$ and $ A \rightarrow \det(A)$. For any symmetric positive
 definite matrix $B$ of order $n$ we have
 \[
  d'(A)B:=  \lim_{t \rightarrow 0}
   \frac{d(A+tB)-d(A)}{t} =\frac{d(A)}{n} \tr(A^{-1}B)~.
 \]
 From the Courant-Fischer eigenvalue characterization we obtain
 $\lambda_i(A^{-1}B) \leq \lambda_i(A^{-1})\|B\|_2 $ for all $i$. Hence
 \[
 \|d'(A)\|_2:= \max_{B~{\rm is~ SPD}} \frac{|d'(A)B|}{\|B\|_2} = \frac{d(A)}{n}
  \max_{B~{\rm is~ SPD}}\frac{ \tr(A^{-1}B)}{\|B\|_2} 
  \leq \frac{d(A)}{n} \tr(A^{-1})~,
 \]
 with equality for $B=I$. Thus for the condition number of the function
 $d$ we have
 \begin{equation}
 \label{cond1}
  \frac{\|A\|_2 \|d'(A)\|_2}{d(A)}=\frac{1}{n} \|A\|_2 \tr(A^{-1}) \leq \kappa(A)~.
\end{equation}
Note that for the diagonal matrix $A={\rm diag}(A_{ii})$ with $A_{11}=1,~
A_{ii}=\alpha \in (0,1)$ for $i>1$, in   the inequality in 
(\ref{cond1}) one obtains equality for $n \rightarrow \infty$.
 For this $A$ and with $\delta A = \varepsilon I$, $\varepsilon > 0$,
 for $n \rightarrow \infty$ we have equality in the second inequality in
  (\ref{basicC}), too. 

For $\tilde{d}(A)=\det(A)=d(A)^n$ the condition number is given by
\begin{equation}
 \label{cond2}
  \frac{\|A\|_2 \|\tilde{d}'(A)\|_2}{\tilde{d}(A)}=
  \frac{\|A\|_2 n d(A)^{n-1}\|d'(A)\|_2}{d(A)^n} =
   \|A\|_2 \tr(A^{-1}) ~,
\end{equation}
i.e. $n$ times larger than the condition number in (\ref{cond1}). The condition
numbers for $d$ and $\tilde{d}$ give an indication of the
sensitivity if the perturbation $\|\delta A\|_2$ is sufficiently small.
Note that the bound in (\ref{basicC}) is  valid for arbitrary symmetric positive definite
perturbations $\delta A$. The bound shows that even for larger
perturbations the function $d(A)$ is well-conditioned at $A$ if $\kappa(A)$ is
not too large. For the function $\tilde{d}(A)$ the effect of
relatively large perturbations can be much worse than for the
asymptotic case ($\delta A \rightarrow 0$),
which is characterized by the condition number in (\ref{cond2}). Consider, for example, for
$0< \varepsilon < \frac{1}{2}$ a perturbation
$\delta A = \varepsilon A$, i.e. $\|\delta A\|_2/\|A\|_2= \varepsilon$. Then
\[
 \frac{\tilde{d}(A+\delta A)- \tilde{d}(A)}{\tilde{d}(A)} = (1+\varepsilon)^n -1
  \geq e^{\frac{1}{2}n \varepsilon}-1~,
 \]
which is very large if, for example, $\varepsilon=10^{-3},~ n=10^5$.

The results in this section show that the numerical approximation
of the function $d(A)$ is a much easier task than the numerical
approximation of $A \rightarrow \det(A)$.
 \section{Sparse approximate inverse}
 \label{SPAI}
 In this section we explain and analyze the construction of a sparse approximate
 inverse of the matrix $A$. Let $A=LL^T$ be the Cholesky factorization of $A$, 
 i.e. $L$ is lower triangular and $L^{-1}AL^{-T}=I$. Note that
 $d(A)=d(L)^{2}=\prod_{i=1}^n L_{ii}^{2/n}$. We will construct a sparse
 lower triangular approximation $G$ of $L^{-1}$ and approximate $d(A)$ by
 $d(G)^{-2}=\prod_{i=1}^n G_{ii}^{-2/n}$. The construction of a sparse approximate
 inverse that we use in this paper was  introduced in 
 \cite{Kaporin,Kol1,Kol2} and can also
 be found in \cite{Axelsson}. Some of the results derived in this section are presented in 
 \cite{Axelsson}, too. 
 
 We first introduce some notation. Let $E \subset \{ (i,j)~|~ 1 \leq i,j \leq n\}$
 be a given sparsity pattern. By $\#E$ we denote the number of elements in $E$. Let
 $S_E$ be the set of $n \times n$ matrices for which all entries are
 set to zero if the corresponding index is {\em not\/} in $E$:
 \[
   S_E=\{ M  \in \Bbb{R}^{n \times n}~|~M_{ij}=0 \mbox{ if } (i,j) \notin E\}~.
\]
For $1 \leq i \leq n$ let $E_i=E \cap \{ (i,j)~|~1 \leq j \leq n\}$. If
$n_i:=\#E_i > 0$ we use the representation
\begin{equation}
  E_i=\{(i,j_1),(i,j_2), \ldots,(i,j_{n_i})\}, \quad 1\leq j_1 < j_2 < \ldots < j_{n_i} \leq n~.
\label{Ei}
\end{equation}
For $n_i > 0$ we define the projection
\begin{equation}
 P_i:~\Bbb{R}^n \rightarrow \Bbb{R}^{n_i}~,\quad P_i(x_1,x_2,\ldots,x_n)^T=(x_{j_1},x_{j_2},
  \ldots,x_{j_{n_i}})^T~.
\label{Pi}
\end{equation}
Note that the matrix 
\[
   P_iAP_i^T:~ \Bbb{R}^{n_i} \rightarrow \Bbb{R}^{n_i}
 \]
 is symmetric positive definite. 
 Typical choices of the sparsity pattern $E$ (cf. \S\ref{Experiments}) are
 such that $n_i$ is a very small number compared to $n$ (e.g. $n_i < 20$).
 In such a case the projected matrix $P_iAP_i^T$ has a 
 small dimension. 
 
 To facilitate the analysis below, we first discuss the construction
 of  an approximate sparse inverse $M_E \in S_E$ in a general framework. For $M_E \in S_E$ we use
 the representation 
 \[
   M_E=  
\left[\begin{array}{c} m_1^T\\
 m_2^T\\  \vdots \\ 
 m_n^T \end{array}\right]~~, \quad m_i \in \Bbb{R}^n~. 
\]   
Note that if $n_i=0$ then $m_i^T=(0,0,\ldots,0)$.

For given $A,B \in \Bbb{R}^{n \times n}$ with $A$ symmetric positive definite we consider
the following problem:
\begin{equation}
\label{Problem}
\mbox{determine } M_E \in S_E \mbox{ such that } (M_EA)_{ij} =B_{ij} \mbox{ for all } (i,j) \in E~.
\end{equation}
In (\ref{Problem}) we have $\#E$ equations to determine $\#E$ entries in $M_E$. We first give two
basic lemmas which will play an important role in the analysis of the sparse approximate inverse
that will be defined in (\ref{apprinverseAB}).
\begin{lemma}
\label{lemma1} The problem (\ref{Problem}) has a unique solution $M_E \in S_E$. If $n_i > 0$ then
the $i$th row of $M_E$ is given by $m_i^T$ with
\begin{equation}
\label{solproblem}
 m_i=P_i^T(P_iAP_i^T)^{-1}P_ib_i~,
 \end{equation}
where $b_i^T$ is the $i$th row of $B$.
\end{lemma}
\begin{proof}
The  equations in  (\ref{Problem}) can be represented as 
\[
  (m_i^TA)_{j_k}=(b_i^T)_{j_k} \mbox{ for all } i \mbox{ with } n_i > 0 \mbox{ and all } k=1,2,\ldots,n_i~,
  \]
  where $m_i^T$ is the $i$th row of $M_E$.
Consider an $i$ with $n_i > 0$. Note that $M_E \in S_E$, hence $P_i^TP_im_i=m_i$. For the unknown entries
in $m_i$ we obtain the system of equations
\[
  (AP_iP_i^Tm_i)_{j_k}=(b_i)_{j_k}~, \quad k=1,2,\ldots,n_i~,
\]
which is equivalent to
\[  P_iAP_i^TP_im_i=P_ib_i ~.
\]
The matrix $P_iAP_i^T$ is symmetric positive definite and thus $m_i$ must satisfy
\[
  P_im_i=(P_iAP_i^T)^{-1}P_ib_i~.
  \]
Using $P_i^TP_im_i=m_i$ we obtain the result in (\ref{solproblem}). The construction in this proof
shows that the solution is unique.
\qquad\end{proof} 

Below we use the Frobenius norm, denoted by $\|\cdot\|_F$:
\begin{equation}
\label{Frobenius}
  \|B\|_F^2=\sum_{i,j=1}^n B_{ij}^2 = \tr(BB^T)~, \quad B \in \Bbb{R}^{n \times n}.
\end{equation}
\begin{lemma}
\label{lemma2}
Let $A=LL^T$ be the Cholesky factorization of $A$ and let $M_E \in S_E$ be the unique solution
of (\ref{Problem}). Then $M_E$ is the unique minimizer of the functional
\begin{equation}
\label{min1}
 M \rightarrow \|(B-MA)L^{-T}\|_F^2=\tr((B-MA)A^{-1}(B-MA)^T),~~ M \in S_E.
\end{equation}
\end{lemma}
\begin{proof}
Let $e_i$ be the  $i$th basis vector in $\Bbb{R}^n$. Take $M \in S_E$. The $i$th rows of 
$M$ and $B$ are denoted by $m_i^T$ and $b_i^T$, respectively. Now note
\begin{equation}
\begin{split}
 \tr((B-MA)A^{-1}(B-MA)^T) &= 
   \sum_{i=1}^n e_i^T(BA^{-1}B^T -MB^T -BM^T+MAM^T)e_i   \\
   &= \tr(BA^{-1}B^T)+\sum_{i=1}^n(-2m_i^Tb_i + m_i^TAm_i)~. 
  \end{split} 
    \label{localmin}
 \end{equation}
 The minimum of the functional (\ref{min1}) is obtained if in (\ref{localmin}) we
 minimize the functionals
 \begin{equation} \label{localmin1}
  m_i \rightarrow -2m_i^Tb_i + m_i^TAm_i~, \quad m_i \in {\mathcal R}(P_i^T)~
\end{equation}
for all $i$ with $n_i > 0$. If we write $m_i=P_i^T\hat{m}_i~,
 \hat{m}_i \in \Bbb{R}^{n_i}$, then
for $n_i > 0$ the functional (\ref{localmin1}) can be rewritten as
\[
  \hat{m}_i \rightarrow -2\hat{m}_i^TP_ib_i + \hat{m}_i^TP_iAP_i^T\hat{m}_i~,\quad \hat{m}_i \in \Bbb{R}^{n_i}.
\]
The unique minimum of this functional is obtained for $\hat{m}_i=(P_iAP_i^T)^{-1}P_ib_i$,
i.e. $m_i=P_i^T(P_iAP_i^T)^{-1}P_ib_i$ for all $i$ with $n_i > 0$. Using Lemma~\ref{lemma1}
it follows that $M_E$  is the unique minimizer of the functional (\ref{min1}).
\qquad\end{proof}
\ \\\\
{\bf Sparse approximate inverse}.
We now introduce the sparse approximate inverse that will be used as an approximation
for $L^{-1}$. For this we choose a lower triangular pattern $E^l \subset \{
(i,j)~|~ 1\leq j \leq i \leq n \}$ and we assume that
$(i,i) \in E^l$ for all $i$. The sparse approximate inverse is constructed in two
steps:
\begin{subequations} \label{apprinverseAB}
\begin{equation} \label{apprinverseA}
 {\rm 1.} \quad \hat{G}_{E^l} \in S_{E^l} \mbox{ such that } (\hat{G}_{E^l}A)_{ij}=\delta_{ij}
    \mbox{ for all } (i,j) \in E^l~,
\end{equation}
\begin{equation} \label{apprinverseB}
 {\rm 2.} \qquad G_{E^l}:= ({\rm diag}(\hat{G}_{E^l}))^{-\frac{1}{2}}\hat{G}_{E^l}~.
\end{equation}
\end{subequations}
The construction of ${G}_{E^l}$ in (\ref{apprinverseAB}) was 
first introduced in \cite{Kaporin}. A theoretical background for this 
factorized sparse inverse is given in \cite{Kol2}. The 
approximate inverse  $\hat{G}_{E^l}$ in (\ref{apprinverseA}) is of the form (\ref{Problem})
with $B=I$. From Lemma~\ref{lemma1} it follows that
in (\ref{apprinverseA}) there is a unique solution $\hat{G}_{E^l}$.
 Note that because $E^l$ is lower triangular and $(i,i) \in E^l$ we have $n_i=\#E^l > 0$ 
for all $i$ and $j_{n_i}=i$ in (\ref{Ei}). Hence it follows from Lemma~\ref{lemma1} that the
$i$th row of $\hat{G}_{E^l}$, denoted by $g_i^T$, is given by
\begin{equation} \label{practice}
\begin{split}
 g_i &= P_i^T(P_iAP_i^T)^{-1}P_ie_i, \qquad i=1,2,\ldots,n,  \\
         &= P_i^T(P_iAP_i^T)^{-1}\hat{e}_i, \quad \mbox{with } ~ \hat{e}_i=(0,\ldots,0,1)^T \in \Bbb{R}^{n_i}~.
   \end{split}
  \end{equation}
 The $i$th entry of $g_i$, i.e. $e_i^Tg_i$, is given by $\hat{e}_i^T(P_iAP_i^T)^{-1}\hat{e}_i$, which
 is strictly positive because $ P_iAP_i^T$ is symmetric positive definite. Hence ${\rm diag}(\hat{G}_{E^l})$
 contains only strictly positive entries and the second 
 step (\ref{apprinverseB}) is well-defined.
 Define $\hat{g}_i=P_ig_i$. The sparse approximate inverse $\hat{G}_{E^l}$ in
 (\ref{apprinverseA}) can be computed by solving the (low dimensional)
 symmetric positive definite systems
 \begin{equation} \label{Practice}
  P_iAP_i^T \hat{g}_i=(0,\ldots,0,1)^T,~~~i=1,2,\ldots,n.
  \end{equation}
  \ \\
We now derive some interesting properties of the sparse approximate
inverse as in (\ref{apprinverseAB}).  
 We start with a minimization property of $\hat{G}_{E^l}$: 
 \begin{theorem}
 \label{theorem1}
 Let $A=LL^T$ be the Cholesky factorization of $A$ and $D:={\rm diag}(L),~\hat{L}:=LD$. 
 $\hat{G}_{E^l}$ as in (\ref{apprinverseA}) is the unique minimizer of the functional
 \begin{equation}
 \label{functheorem1}
 G \rightarrow \|(I-G\hat{L})D^{-1}\|_F^2 = \tr((I-G\hat{L})D^{-2}(I-G\hat{L})^T),~~ G \in S_{E^l}.
 \end{equation}
 \end{theorem}
 \begin{proof}
 The construction of $\hat{G}_{E^l}$ in (\ref{apprinverseA}) is as in (\ref{Problem}) with
 $E=E^l$, $B=I$. Hence Lemma~\ref{lemma2} is applicable with $B=I$. It follows that
 $\hat{G}_{E^l}$ is the unique minimizer of
 \begin{equation}
  G \rightarrow \|(I-GA)L^{-T}\|_F^2~, \quad G \in S_{E^l}~.
\label{loc1}
\end{equation}
Decompose $L^{-T}$ as $L^{-T}=D^{-1}+R$ with $R$ strictly upper triangular.  Then 
$D^{-1}-GL$ and $R$ are lower and strictly upper triangular, respectively, and we obtain:
\begin{eqnarray*}
\|(I-GA)L^{-T}\|_F^2 &=& \|(I-GLL^T)L^{-T}\|_F^2 = \|D^{-1} +R -GL \|_F^2 \\
  &=& \|D^{-1} -GL\|_F^2 +\|R\|_F^2 = \|(I-G\hat{L})D^{-1}\|_F^2 +\|R\|_F^2~.
\end{eqnarray*}
Hence the minimizers in (\ref{loc1}) and (\ref{functheorem1}) are the same.
 \qquad\end{proof}
 ~ \\
\begin{remark}{\rm
From the result in Theorem~\ref{theorem1} we see that in a scaled Frobenius norm
(scaling with $D^{-1}$) $\hat{G}_{E^l}$ is the optimal approximation of $\hat{L}^{-1}$
in the set $S_{E^l}$, in the sense that $\hat{G}_{E^l}\hat{L}$ is closest to the identity.
A seemingly more natural minimization problem is
\begin{equation}
\label{loc2}
 \min_{G \in S_{E^l}} \|I -GL\|_F ~,
 \end{equation}
i.e. we directly approximate $L^{-1}$ (instead of $\hat{L}^{-1}$) and do not use the scaling
with $D^{-1}$. The minimization problem (\ref{loc2}) is of the form as in Lemma~\ref{lemma2}
with $B=L^T$, $E=E^l$. Hence the unique minimizer in (\ref{loc2}), denoted by
 $\tilde{G}_{E^l}$, must satisfy (\ref{Problem})
with $B=L^T$:
\begin{equation}
 \label{loc3}
 (\tilde{G}_{E^l}A)_{ij}=L_{ji} \quad \mbox{for all } (i,j) \in E^l~.
 \end{equation}
Because $E^l$ contains only indices $(i,j)$ with $i \geq j$ and $L_{ji}=0$ for $i > j$ it follows
that $\tilde{G}_{E^l} \in S_{E^l}$ must satisfy
\begin{equation}
\label{loc4}
 (\tilde{G}_{E^l}A)_{ij} = 
  \left\{ \begin{array}{l}
    0 ~\mbox{ if } i \neq j \\
    L_{ii}~ \mbox{ if } i=j    \end{array} \right. \qquad \mbox{ for all } (i,j) \in E^l~.
\end{equation} 
This is similar to the system of equations in (\ref{apprinverseA}), which characterizes 
$\hat{G}_{E^l}$. However, in (\ref{loc4}) one needs the values $L_{ii}$, which 
in general  are not available. Hence opposite to the minimization problem related to
the functional (\ref{functheorem1}) the 
 minimization problem (\ref{loc2}) is in general not solvable with acceptable computational costs.
}
 \qquad  $\Box$ \end{remark}

The following lemma will be used in the proof of Theorem~\ref{mainthm}.
\begin{lemma} \label{alfa}
Let $\hat{G}_{E^l}$ be as in (\ref{apprinverseA}). Decompose $\hat{G}_{E^l}$ as
$\hat{G}_{E^l}=\hat{D}(I-\hat{L})$, with $\hat{D}$ diagonal and $\hat{L}$ strictly
lower triangular. Define $E_-^l:=E^l \setminus \{ (i,i)~|~ 1 \leq i \leq n \}$.
Then $\hat{L}$ is the unique minimizer of the functional
\begin{equation}
  L \rightarrow \tr((I-L)A(I-L^T))~, \qquad L \in S_{E_-^l}~,
  \label{13}
\end{equation}
and also of the functional
\begin{equation}
 L \rightarrow \det[{\rm diag}((I-L)A(I-L^T))]~, \qquad L \in S_{E_-^l} ~.
 \label{14}
\end{equation}
Furthermore, for $\hat{D}$ we have
\begin{equation}
 \hat{D}=[{\rm diag}((I-\hat{L})A(I-\hat{L}^T))]^{-1}~.
 \label{15}
 \end{equation}
 \end{lemma}
\begin{proof}
From the construction in (\ref{apprinverseA}) it follows that
\[
  ((I-\hat{L})A)_{ij}=0 ~~\mbox{ for all } (i,j)  \in E_-^l~,
  \]
i.e., $\hat{L} \in S_{E_-^l}$ is such that $(\hat{L}A)_{ij}=A_{ij}$ for all
$(i,j) \in S_{E_-^l}$. This is of the form (\ref{Problem}) with $B=A$, $E=E_-^l$.
From Lemma~\ref{lemma2} we obtain that $\hat{L}$ is the unique minimizer of the
functional
\[
 L \rightarrow \tr((A-LA)A^{-1}(A-LA)^T)= \tr((I-L)A(I-L^T))~, \quad L \in S_{E_-^l}~,
 \]
 i.e., of the functional (\ref{13}).
 From the proof of Lemma~\ref{lemma2}, with $B=A$, it follows that the 
 minimization problem
 \[
  \min_{L \in S_{E_-^l}} \tr((I-L)A(I-L^T)) 
  \]
decouples into seperate minimization problems (cf. (\ref{localmin1})) for the rows
of $L$:
\begin{equation}
 \min_{l_i \in \mathcal{R}(P_i^T)} \{ -2l_i^T a_i + l_i^T Al_i \}
 \label{16}
\end{equation}
 for all $i$ with $n_i > 0$. Here $l_i^T$ and $a_i^T$ are the $i$th rows of $L$ and
 $A$, respectively. The minimization problem corresponding to (\ref{14}) is
 \[
   \min_{L \in S_{E_-^l}} \prod_{i=1}^n ((I-L)A(I-L^T))_{ii} =
     \min_{L \in S_{E_-^l}} \prod_{i=1}^n (A_{ii} -2 l_i^Ta_i +l_i^TAl_i) ~.
 \]
 This decouples into the same minimization problems as in (\ref{16}). Hence the functionals
 in (\ref{13}) and (\ref{14}) have the same minimizer.
 
 Let $J={\rm diag}((I-\hat{L})A(I-\hat{L}^T))$. Using the construction of
 $\hat{G}_{E^l}$ in (\ref{apprinverseA}) we obtain
 \begin{eqnarray*}
  \hat{D}_{ii}^2 J_{ii}  &=&
   (\hat{D}(I-\hat{L})A(I-\hat{L}^T)\hat{D})_{ii} =
   (\hat{G}_{E^l}A\hat{G}_{E^l}^T)_{ii} \\
    &=& \sum_{k=1}^n (\hat{G}_{E^l}A)_{ik}(\hat{G}_{E^l})_{ik} =
    \sum \begin{Sb}
        k=1 \\
       (i,k) \in E^l 
       \end{Sb} ^n \delta_{ik}(\hat{G}_{E^l})_{ik} \\
    ~   &=& (\hat{G}_{E^l})_{ii} = \hat{D}_{ii} ~.
 \end{eqnarray*}
Hence $\hat{D}_{ii}=J_{ii}^{-1}$ holds for all $i$, i.e., (\ref{15}) holds.
\qquad\end{proof}
\begin{corollary}{\rm From (\ref{15}) it follows that ${\rm diag}(\hat{G}_{E^l}
A\hat{G}_{E^l})={\rm diag}(\hat{G}_{E^l})$  holds and thus, using (\ref{apprinverseB})
we obtain
\begin{equation}
 {\rm diag}(G_{E^l}
AG_{E^l})=I
\label{GAGisI}
\end{equation}
for  the sparse approximate inverse $G_{E^l}$.
} $~~ \qquad \qquad \Box$  \end{corollary}

The following theorem gives a main result in the theory of approximate 
inverses. It was first derived in
\cite{Kol2}. A proof can be found in \cite{Axelsson}, too.
\begin{theorem}
Let $G_{E^l}$ be the approximate inverse in (\ref{apprinverseAB}). Then $G_{E^l}$ is the
unique minimizer of the functional
\begin{equation}
 G \rightarrow  \frac{\frac{1}{n} \tr(GAG^T)}{ \det(GAG^T)^{\frac{1}{n}}}~,
\qquad G \in S_{E^l} ~.
 \label{mincond}
 \end{equation} 
 \label{mainthm}
\end{theorem} 
\begin{proof}
For $G \in S_{E^l}$ we use the decomposition $G=D(I-L)$, with $D$ diagonal and $L \in S_{E_-^l}$.
Furthermore, for $L \in S_{E_-^l}$, $J_L:={\rm diag}((I-L)A(I-L^T))$. Now note \\
 \begin{equation}
 \begin{split}
 \frac{\frac{1}{n} \tr(GAG^T)}{ \det(GAG^T)^{\frac{1}{n}}}
 &= \det(A)^{-\frac{1}{n}} \frac{ \frac{1}{n} \tr((D(I-L)A(I-L^T)D)}{ \det({G^2})^{\frac{1}{n}}} 
   = \det(A)^{-\frac{1}{n}} \frac{\frac{1}{n} \tr(D^2J_L)}{ \det(D^2)^{\frac{1}{n}}}  \\
   &= \det(A)^{-\frac{1}{n}}  \frac{\frac{1}{n} \tr(D^2J_L)}{ \det(D^2 J_L)^{\frac{1}{n}}} \det(J_L)^{\frac{1}{n}} 
    \geq \det(A)^{-\frac{1}{n}} \det(J_L)^{\frac{1}{n}} ~.
    \label{loc78}
    \end{split}
\end{equation}
The  inequality in (\ref{loc78}) follows from the inequality between the arithmetic and geometric mean:
$\frac{1}{n} \sum_{i=1}^n \alpha_i \geq (\prod_{i=1}^n \alpha_i)^{1/n}$ for $\alpha_i \geq 0$.

For  $\hat{G}_{E^l}$ in (\ref{apprinverseA}) we use the decomposition $\hat{G}_{E^l}=\hat{D}(I-\hat{L})$.
For the approximate inverse $G_{E^l}$ we then have $G_{E^l}= ({\rm diag}(\hat{G}_{E^l}))^{-\frac{1}{2}
} \hat{G}_{E^l}=\hat{D}^{\frac{1}{2}}(I-\hat{L})$.
From Lemma~\ref{alfa} (\ref{14}) it follows that $\det(J_L) \geq \det(J_{\hat{L}})$ for all
$L \in S_{E_-^l}$. Furthermore from Lemma~\ref{alfa}
(\ref{15}) we obtain that for $G_{E^l}=\hat{D}^{\frac{1}{2}}(I-\hat{L})$ we have
$(\hat{D}^{\frac{1}{2}})^2J_{\hat{L}} =I$ and thus equality in (\ref{loc78}) for $G=G_{E^l}$
. We conclude that 
$G_{E^l}$ is the unique minimizer of the functional in (\ref{mincond}).
\qquad\end{proof}
~ \\ \\
\begin{remark} \label{KA}
{\rm The quantity
\[
 K(A)= \frac{\frac{1}{n} \tr(A)}{\det(A)^{\frac{1}{n}}} 
 \]
 can be seen as a nonstandard condition number 
 (cf. \cite{Axelsson,Kaporin}). Properties
 of this quantity are given in \cite{Axelsson} (Theorem 13.5). One elementary property is
 \[
   1 \leq K(A) \leq \frac{\lambda_n}{\lambda_1} = \kappa(A) ~. \qquad \qquad \Box
   \]
}  \end{remark}
  \begin{corollary} \label{gevolg} {\rm
  For the approximate inverse $G_{E^l}$ as in (\ref{apprinverseAB}) we have 
  (cf.(\ref{GAGisI}))   
 \[
    1 \leq K(G_{E^l}AG_{E^l}^T) = \frac{1}{ \det(G_{E^l}AG_{E^l}^T)^{\frac{1}{n}}}~,
\]
i.e.,
\begin{equation}
\label{upperbound}
 d(A) \leq \det(G_{E^l}^2)^{-\frac{1}{n}} = \prod_{i=1}^n (G_{E^l})_{ii}^{-\frac{2}{n}}
  = \prod_{i=1}^n (\hat{G}_{E^l})_{ii}^{-\frac{1}{n}}~.
\end{equation}
Let $\tilde{E}^l$ be a lower triangular sparsity pattern that is larger than $E^l$,
i.e., $E^l \subset \tilde{E}^l \subset \{ (i,j)~|~ 1 \leq j \leq i \leq n \}$. From
the optimality result in Theorem~\ref{mainthm} it follows that
\begin{equation}
\label{improvement}
 1 \leq K(G_{\tilde{E}^l}AG_{\tilde{E}^l}^T) \leq K(G_{E^l}AG_{E^l}^T)~. 
\end{equation}
\hspace*{10cm} $ \qquad \qquad \Box$ 
} \end{corollary}
Motivated by the theoretical results in
Corollary~\ref{gevolg}   we {\em propose to use the  sparse
 approximate inverse\/}
$G_{E^l}$ as in
(\ref{apprinverseAB}) for approximating $d(A)$: Take 
$d(G_{E^l})^{-2}=d(\hat{G}_{E^l})^{-1}$ 
as an estimate for $d(A)$.
Some properties of this method are discussed in the following remark. 
\begin{remark} \label{properties} {\rm
We consider the method of approximating $d(A)$ by
 $d(G_{E^l})^{-2}=d(\hat{G}_{E^l})^{-1}$. The practical realization of this method
 boils down to chosing a sparsity pattern $E^l$ and solving the (small) 
 systems in (\ref{Practice}).  
We list a few properties of this approach:
 \begin{remunerate}
\item The sparse approximate inverse exists for every symmetric positive
 definite $A$. Note that such an existence result does not hold for the incomplete
  Cholesky factorization. Furthermore, this factorization is obtained by 
  solving low dimensional symmetric positive definite systems
 of the form $P_iAP_i^T\hat{g}_i=\hat{e}_i$ (cf. (\ref{Practice})), which
 can be realized in a stable way. 
\item The systems $P_iAP_i^T\hat{g}_i=\hat{e}_i$, $i=1,2,\dots,n$, can be solved in
parallel.
\item For the computation of  $d(G_{E^l})^{-2}=d(\hat{G}_{E^l})^{-1}$ we only need the diagonal entries
of $\hat{G}_{E^l}$ (cf. (\ref{upperbound})). In the systems $P_iAP_i^T\hat{g}_i=\hat{e}_i$ we then only 
have to compute the
last entry of $\hat{g}_i$, i.e. $(\hat{g}_i)_{n_i}$. If these systems are solved using
the Cholesky factorization, $P_iAP_i^T=:L_iL_i^T$ ($L_i$ lower triangular) we only need the
$(n_i,n_i)$ entry of $L_i$, since  $(\hat{g}_i)_{n_i}=(L_i)_{n_in_i}^{-2}$.
\item The sparse approximate inverse has an optimality property related to the determinant:
 The functional $G \rightarrow K(GAG^T)~,~ G \in S_{E^l}$, is minimal for
  $G_{E^l}$. From this the inequality (\ref{upperbound}) and the monotonicity result
  (\ref{improvement}) follow.
\item  From(\ref{upperbound}) we obtain the upper bound $0$ for the relative error
 $d(A)/d(G_{E^l})^{-2} -1 $. In \S\ref{ErrorEstimation} we will derive useful 
 lower bounds for this relative error. These are a posteriori error bounds 
 which use the matrix $G_{E^l}$.  \qquad $\Box$
\end{remunerate}
}  \end{remark}
 \section{A posteriori error estimation}
 \label{ErrorEstimation}
 In the previous section it has been explained how an estimate $d(G_{E^l})^{-2}$ of $d(A)$
 can be computed. From (\ref{upperbound}) we have the error bound
 \begin{equation}
   \frac{d(A)}{d(G_{E^l})^{-2}} \leq 1~.
  \label{upper}
\end{equation}
In this section we will discuss a posteriori  estimators 
for the error  $d(A)/d(G_{E^l})^{-2}$. In \S\ref{method1}
we apply the analysis from \cite{Bai} 
to derive  an a posteriori lower bound for the quantity in (\ref{upper}). 
This approach results in safe, but often rather pessimistic bounds for
the error. In \S\ref{method2} we propose a very simple
stochastic method  for error estimation. This method,
although it does not yield guaranteed bounds for the error,
turns out to be very useful in practice.
\subsection{Error estimation based on bounds from \cite{Bai} }
\label{method1}
In this section we show how the analysis from \cite{Bai} 
 can be used to obtain an error estimator. 
We first recall a main result from \cite{Bai} (Theorem 2). Let $A$ be a symmetric positive matrix of
order $n$, $\mu_1=\tr(A),~\mu_2=\|A\|_F^2$ and $ \sigma(A) \subset [\alpha, \beta]$ with
$\alpha > 0$. Then:
\begin{equation}
\begin{split}
\label{Golub}
 ~ & \exp \left( \frac{1}{n} [\ln \alpha ~~ \ln t_l] 
\left[ \begin{array}{cc} \alpha & t_l \\
                        \alpha^2 & t_l^2
       \end{array} \right] ^{-1} 
       \left[ \begin{array}{c} \mu_1 \\ \mu_2 \end{array} \right] \right) 
       \leq d(A)   \leq  \\
 ~ & \exp \left( \frac{1}{n} [\ln \beta ~~ \ln t_u] 
\left[ \begin{array}{cc} \beta & t_u \\
                        \beta^2 & t_u^2
       \end{array} \right] ^{-1} 
       \left[ \begin{array}{c} \mu_1 \\ \mu_2 \end{array} \right] \right) ~, 
 \end{split}
 \end{equation}
 where $t_l=\frac{\alpha \mu_1 -\mu_2}{\alpha n - \mu_1},~~t_u=\frac{\beta \mu_1 -\mu_2}{\beta n - \mu_1}$.
 
 In \cite{Bai} this result is applied to obtain computable bounds for $d(A)$. Often these bounds yield
 rather poor estimates of $d(A)$. In the present paper we approximate $d(A)$ by
 $d(G_{E^l})^{-2}$ and use the result in (\ref{Golub}) for {\em error estimation\/}.
 The upper bound (\ref{upper}) turns out to be satisfactory in numerical experiments
 (cf. \S\ref{Experiments}). Therefore we restrict ourselves to the derivation
 of a lower bound for $d(A)/d(G_{E^l})^{-2}$, based on the left inequality in (\ref{Golub}).
 \begin{theorem} \label{Golubhere}
 Let $G_{E^l}$ be the approximate inverse from (\ref{apprinverseAB}) and
 $0 < \alpha \leq \lambda_{\min}(G_{E^l}AG_{E^l}^T)$, $\mu:=\frac{1}{n} \|G_{E^l}AG_{E^l}^T\|_F^2$,
 $\delta:= \mu -1$. The following holds: $\alpha \leq 1$, $\delta \geq 0$ and
 \begin{equation}
 \label{errorbounds}
 \exp \left( \frac{1}{(\alpha-1)^2 + \delta} \left( \delta \ln \alpha +
   (1- \alpha)^2 \ln (1+ \frac{\delta}{1 - \alpha} ) \right) \right) 
   \leq \frac{d(A)}{d(G_{E^l})^{-2} } \leq 1~.
\end{equation}
 \end{theorem}
\begin{proof}
The right inequality in (\ref{errorbounds}) is already given in (\ref{upper}).
We introduce the notation $ \tau_1 \leq \tau_2 \leq \ldots \leq \tau_n $
for the eigenvalues of $G_{E^l}AG_{E^l}^T$. From (\ref{GAGisI}) we obtain 
 $\frac{1}{n} \sum_{i=1}^n \tau_i =1$ and
from this it follows that $\alpha \leq \tau_1 \leq 1$ holds. Furthermore,
\[
 1= (\frac{1}{n} \sum_{i=1}^n \tau_i)^2 \leq \frac{1}{n}  \sum_{i=1}^n \tau_i^2 
 \]
 yields $\mu = \frac{1}{n} \tr((G_{E^l}AG_{E^l}^T)^2) \geq 1$ and thus $\delta \geq 0$.
 We now use the left inequality in (\ref{Golub}) applied to the
 matrix $G_{E^l}AG_{E^l}^T$. Note that 
 \[
  \mu_1=\tr(G_{E^l}AG_{E^l}^T)=n,~~\mu_2=n\mu,~~t_l=\frac{\alpha \mu_1 -\mu_2}{\alpha n -\mu_1}=
   1+ \frac{\delta}{1 - \alpha}~.
  \]
 A simple computation yields
 \begin{equation}
 \label{loc89}
  \frac{1}{n} \left[ \begin{array}{cc} \alpha & t_l \\
                        \alpha^2 & t_l^2
       \end{array} \right] ^{-1} 
       \left[ \begin{array}{c} \mu_1 \\ \mu_2 \end{array} \right]
       = \frac{1}{t_l - \alpha}
        \left[ \begin{array}{c} \frac{\delta}{1 - \alpha} \\  1-\alpha \end{array} \right]~,
\end{equation}
and
\begin{equation}
\label{loc90}
 t_l- \alpha = \frac{(1-\alpha)^2 + \delta}{1 - \alpha}~.
 \end{equation}
Substitution of (\ref{loc90}) in (\ref{loc89}) results in
\begin{eqnarray*}
 \frac{1}{n} [\ln \alpha ~~ \ln t_l] 
\left[ \begin{array}{cc} \alpha & t_l \\
                        \alpha^2 & t_l^2
       \end{array} \right] ^{-1} 
       \left[ \begin{array}{c} \mu_1 \\ \mu_2 \end{array} \right] &=& 
       \frac{1}{(1- \alpha)^2 + \delta} \left( \delta \ln \alpha + (1 - \alpha)^2 \ln t_l \right) \\
       &=&  \frac{1}{(1- \alpha)^2 + \delta} \left( \delta \ln \alpha + (1 - \alpha)^2 \ln (1+
        \frac{\delta}{1 -\alpha}) \right)~.
\end{eqnarray*}
Using this the left inequality in (\ref{errorbounds}) follows from the left inequality in 
(\ref{Golub}). 
 \qquad\end{proof}     
\\\\\ 
Note that for the lower bound in (\ref{errorbounds}) to be computable, we need 
 $\mu=\frac{1}{n}\|G_{E^l}AG_{E^l}^T\|_F^2$ and a strictly positive lower bound
 $\alpha$ for  the smallest eigenvalue of $G_{E^l}AG_{E^l}^T$. We now discuss
 methods for computing $\alpha$ and $\mu$. These methods are used in the
 numerical experiments in \S\ref{Experiments}.\\\\
 We first discuss two methods for computing $\alpha$. The first method, which
 can be applied if $A$ is an $M$-matrix, is
 based on the following lemma, where we use the notation ${\bf 1} = (1,1,\ldots,1)^T \in \Bbb{R}^n$.
 \begin{lemma} \label{Mmatrix}
 Let $A$ be a symmetric positive definite matrix of order $n$ with $A_{ij} \leq 0$ for
 all $i \neq j$ and $G_{E^l}$ its sparse approximate inverse (\ref{apprinverseAB}). 
 Furthermore, let $z$ be such that 
 \[
    \|G_{E^l}AG_{E^l}^T z - {\bf 1} \|_{\infty} \leq \eta < 1~.
\]
Then 
\begin{equation} \label{Mlower}
    (1-\eta) \|z\|_{\infty}^{-1} \leq \lambda_{\min}(G_{E^l}AG_{E^l}^T) 
\end{equation}
holds.
  \end{lemma}
 \begin{proof}
 From the assumptions it follows that $A$ is an $M$-matrix. In 
 \cite{Kol2} (Theorem 4.1) it is
 proved that then $G_{E^l}AG_{E^l}^T$ is an $M$-matrix, too. Let $z^{\ast}=(G_{E^l}AG_{E^l}^T)^{-1}{\bf 1}$.
 Because $(G_{E^l}AG_{E^l}^T)^{-1}$ has only nonnegative entries it follows that
\begin{eqnarray*}
 \|(G_{E^l}AG_{E^l}^T)^{-1}\|_{\infty} &=& \|z^{\ast}\|_{\infty}=\|z + (z^{\ast}-z)\|_{\infty} \\
  &\leq & \|z\|_{\infty} + \|(G_{E^l}AG_{E^l}^T)^{-1}\|_{\infty} \|G_{E^l}AG_{E^l}^Tz - {\bf 1}\|_{\infty}
  \\
  & \leq & \|z\|_{\infty} + \|(G_{E^l}AG_{E^l}^T)^{-1}\|_{\infty} \eta ~.
\end{eqnarray*}
Hence $ \|(G_{E^l}AG_{E^l}^T)^{-1}\|_{\infty}^{-1} \geq (1 - \eta) \|z\|_{\infty}^{-1} $. Using
 $\lambda_{\min}(G_{E^l}AG_{E^l}^T) \geq  \|((G_{E^l}AG_{E^l}^T)^{-1}\|_{\infty}^{-1}$ we
 obtain the result (\ref{Mlower}).
 \qquad\end{proof}

Based on this lemma we obtain the following method for computing $\alpha$. Choose $0 < \eta \ll 1 $
and apply the conjugate gradient method to the system $G_{E^l}AG_{E^l}^Tz^{\ast}={\bf 1}$. This
results in approximations $z^0,z^1,\dots $ of $z^{\ast}$. One iterates until the
stopping criterion $d^j:=\|G_{E^l}AG_{E^l}^Tz^j - {\bf 1}\|_{\infty} \leq \eta $ is
satisfied. Then take $\alpha:= (1-d^j)\|z^j\|_{\infty}^{-1}$. In view of efficiency one
should not take a very small tolerance $\eta$. In our experiments in \S\ref{Experiments} we
use $\eta=0.2$ and $z^0={\bf 1}$. 
Note that the CG method is applied to a system with
the {\em preconditioned\/} matrix $G_{E^l}AG_{E^l}^T$. In
situations where the preconditioning is
effective   one may expect that
relatively few CG iterations are needed to compute $z^j$ such that
$\|G_{E^l}AG_{E^l}^Tz^j - {\bf 1}\|_{\infty} \leq \eta $ is satisfied.
Results of numerical  experiments are presented in \S\ref{Experiments}.
\\\\
As a second method for determining $\alpha$, which is applicable to any symmetric positive
definite $A$, we propose the Lanczos method for approximating eigenvalues applied to
the matrix $G_{E^l}AG_{E^l}^T$. This method yields a decreasing sequence
$\lambda_1^{(1)} \geq \lambda_1^{(2)} \geq \dots \geq \lambda_1^{(j)} \geq \lambda_{\min}(G_{E^l}AG_{E^l}^T)$
of approximations $\lambda_1^{(j)}$ of $\lambda_{\min}(G_{E^l}AG_{E^l}^T)$. If
\begin{equation}
\label{errorLanczos}
 \lambda_1^{(j)} - \lambda_{\min}(G_{E^l}AG_{E^l}^T) < \varepsilon
\end{equation}
holds, then $\alpha=\lambda_1^{(j)} - \varepsilon $ can be used in Theorem~\ref{Golubhere}. However,
in practice it is usually not known how to obtain reasonable values for 
$\varepsilon$ in (\ref{errorLanczos}). Therefore,
in our experiments we use a simple heuristic for error estimation (instead of a rigorous bound
as in (\ref{errorLanczos})), based on the observed convergence behaviour
 of $\lambda_1^{(j)}$ (cf. \S\ref{Experiments}).
 \\
 It is known that for the
Lanczos method the convergence to extreme eigenvalues is relatively fast.
Moreover, it often occurs
that the small eigenvalues of the preconditioned matrix $G_{E^l}AG_{E^l}^T$ are well-separated
from the rest of the spectrum, which has a positive effect on the convergence speed  
$\lambda_1^{(j)} \rightarrow \lambda_{\min}(G_{E^l}AG_{E^l}^T)$.
 In numerical experiments we
indeed observe that often already after a few Lanczos iterations we have an approximation 
of $\lambda_{\min}(G_{E^l}AG_{E^l}^T)$ with an estimated  relative error of
 a few percent.
However, for the $\alpha$ computed in this second method we do not have a rigorous analysis which
guarantees that $0 < \alpha < \lambda_{\min}(G_{E^l}AG_{E^l}^T)$ holds.
 Nevertheless, from
numerical experiments we see that this  method is satisfactory. This is partly
explained by the relatively fast convergence of the Lanczos method towards the
smallest eigenvalue. A further explanation  follows from the form of the
lower bound in (\ref{errorbounds}). For $\alpha \ll 1, ~ \delta \ll 1$, which is
typically the case in our experiments in \S\ref{Experiments}, this lower bound 
essentially behaves like $\exp(\delta \ln \alpha)=:g(\alpha)$. 
Note that
$0 < \frac{g'(\alpha) \alpha}{g(\alpha)} = \delta \ll 1$ holds. Hence the sensitivity
of the lower bound with respect to perturbations in $\alpha$ is very mild.  \\\\
We now discuss the computation of the quantity 
$\mu= \frac{1}{n} \|G_{E^l}AG_{E^l}^T\|_F^2$, which is needed
in (\ref{errorbounds}). Clearly, for computing $\mu$ one needs the matrices $G_{E^l}$
and $A$. To avoid unnecessary storage requirements one should not compute the matrix
$X:=G_{E^l}AG_{E^l}^T$ and then determine $\mu = \frac{1}{n}\|X\|_F^2$. A with respect to storage
more efficient approach can be based on 
\[
  \|G_{E^l}AG_{E^l}^T\|_F^2= \sum_{i=1}^n \|G_{E^l}AG_{E^l}^T e_i\|_2^2~,
  \]
where $e_i$ is the $i$th basis vector in $\Bbb{R}^n$. For the computation of
$\|G_{E^l}AG_{E^l}^T e_i\|_2^2$, $i=1,2,\ldots,n$, which can be done in parallel,
one needs only sparse matrix-vector multiplications with the matrices 
$G_{E^l}$ and $A$. Furthermore, for the computation of
$AG_{E^l}^Te_i$ one can use that
$(DG_{E^l}A)_{ij}=(\hat{G}_{E^l}A)_{ij}=\delta_{ij}$ for $(i,j) \in E^l$,
with $D:={\rm diag}(G_{E^l})$ (cf. (\ref{apprinverseAB})). It follows
from (\ref{practice}) that
\begin{eqnarray*}
  AG_{E^l}^Te_i &=& (I-P_i^TP_i)AG_{E^l}^Te_i + P_i^TP_i AG_{E^l}^Te_i \\
   &=& (I-P_i^TP_i)AG_{E^l}^Te_i + P_i^TP_i A\hat{G}_{E^l}^T D^{-1}e_i \\
   &=& (I-P_i^TP_i)AG_{E^l}^Te_i + D_{ii}^{-1} e_i 
\end{eqnarray*}
holds.
\begin{remark}
\label{opmerror}
{\rm Note that for the error estimators discussed in this section
the matrix $G_{E^l}$ must be available (and thus stored), whereas
for the computation of the approximation
$d(G_{E^l})^{-2}$ of $d(A)$ we do not have to store the matrix
$G_{E^l}$ (cf. Remark~\ref{properties} item 3). Furthermore, as we
will
see in \S\ref{Experiments}, the computation of these error estimators
is relatively expensive.
} \qquad $\Box$ \end{remark}

\subsection{Error estimation based on a Monte Carlo approach}
\label{method2}
In this section we discuss a simple error estimation method which
turns out to be  useful in practice. Opposite to those treated
in the previous section this method does not yield (an approximation of)
bounds for the error. \\
The exact error is given by 
\[
 \frac{d(A)}{d(G_{E^l})^{-2}}=d(G_{E^l}AG_{E^l}^T)=d({\mathcal E}_{E^l})~,
\]
where
$
{\mathcal E}_{E^l}:=G_{E^l}AG_{E^l}^T 
$
is a sparse symmetric positive definite
matrix. Fore ease of presentation we assume that the pattern $E^l$ is
sufficiently
large such that
\begin{equation}
\label{rho}
 \rho ( I - {\mathcal E}_{E^l}) < 1
\end{equation}
holds. In \cite{Kol2} it is proved that if $A$ is an $M$-matrix  or a
(block) $H$-matrix then (\ref{rho}) is satisfied for every 
lower triangular pattern $E^l$. In the numerical experiments (cf.
\S\ref{Experiments}) with matrices which are not $M$-matrices or
(block) $H$-matrices (\ref{rho}) turns out to be satisfied for
standard choices of $E^l$. We note that if (\ref{rho}) does not hold then
the technique discussed below can still be applied if one replaces
${\mathcal E}_{E^l}$ by $\omega {\mathcal E}_{E^l}$ with  $\omega > 0$ a
suitable damping factor such that   $\rho ( I - \omega {\mathcal E}_{E^l}) < 1
$ is satisfied.

For the exact error we obtain, using a Taylor expansion of
$\ln(I-B)$ for $B \in \Bbb{R}^{n \times n}$ with $\rho(B) < 1$ (cf. \cite{Golub-VanLoan}):
\begin{equation}
\begin{split}
 d({\mathcal E}_{E^l})&= \exp \left( \frac{1}{n} \ln ( \det ({\mathcal E}_{E^l}))
   \right) = \exp \left( \frac{1}{n} \tr ( \ln ({\mathcal E}_{E^l} )) \right) 
     \\
   &= \exp \left( \frac{1}{n} \tr ( \ln (I- (I-{\mathcal E}_{E^l}))) \right)
    = \exp \left( - \frac{1}{n} \tr( \sum_{k=1}^{\infty} 
    \frac{ (I-{\mathcal E}_{E^l})^k}{k}) \right) 
   \end{split}
    \label{expansion}
\end{equation}
Hence,  an error estimation can be based on estimates for the
partial sums 
$S_m:= \sum_{k=1}^m \frac{1}{k} \tr((I-{\mathcal E}_{E^l})^k)$. The construction
of $G_{E^l}$ is such that ${\rm diag}({\mathcal E}_{E^l})=I$
(cf. (\ref{GAGisI})) and thus
$\tr({\mathcal E}_{E^l})=n$ and $S_1=0$. For $S_2$ we have
\begin{equation} \label{S2}
 S_2=\frac{1}{2} \tr ((I-{\mathcal E}_{E^l})^2) = \frac{1}{2} \tr(I-
  2{\mathcal E}_{E^l} +{\mathcal E}_{E^l}^2) = -\frac{1}{2}n +
   \frac{1}{2} \tr ({\mathcal E}_{E^l}^2)~.
\end{equation}
For $S_3$ we obtain
\begin{equation} \label{S3}
S_3=\frac{1}{2} \tr((I-{\mathcal E}_{E^l})^2)+ \frac{1}{3} 
\tr((I-{\mathcal E}_{E^l})^3)
 = - \frac{7}{6} n + \frac{3}{2} \tr({\mathcal E}_{E^l}^2)- 
 \frac{1}{3}\tr({\mathcal E}_{E^l}^3)~.
 \end{equation}
 Note that in $S_2$ and $S_3$ the quantity 
 $\tr({\mathcal E}_{E^l}^2)=\|{\mathcal E}_{E^l}
 \|_F^2 = \|G_{E^l}AG_{E^l}^T\|_F^2$ occurs  which is also used in the error
 estimator in \S\ref{method1}. In this section we use a {\em Monte Carlo method\/}
  to approximate
 the trace quantities in $S_m$.
 The method we use is based on the following proposition 
 \cite{Hutchinson,Bai2}.
\begin{proposition}\label{Carlo}
Let $H$ be a symmetric matrix of order $n$ with $\tr(H) \neq 0$.  Let $V$ be
the discrete random variable which takes the values 1 and $-1$ each with
probability $0.5$ and let $z$ be a vector of $n$ independent samples from
$V$. Then $z^THz$ is an unbiased estimator of $\tr(H)$:
\[
 E(z^THz) = \tr(H)~,
\]
and
\[
  var(z^THz)=2 \sum_{i \neq j} h_{ij}^2~.
 \] 
\end{proposition}
For approximating the trace quantity in $S_2$ we use the
 following Monte Carlo algorithm:
  \\\\
 \hspace*{0.5cm} for $j=1,2,\dots,M $ \\
 \hspace*{1cm} 1.~ Generate $z_j \in \Bbb{R}^n$ with entries which are uniformly
 distributed in $(0,1)$.\\
 \hspace*{1cm} 2.~  If $(z_j)_i < 0.5 $ then $(z_j)_i:=-1$, otherwise, $(z_j)_i:=1$.
 \\
 \hspace*{1cm} 3.~  $y_j:={\mathcal E}_{E^l}z_j$, \quad
  $\alpha_j:=y_j^Ty_j$.\\\\
Based on Proposition~\ref{Carlo} and (\ref{S2}) we use 
 \begin{equation} \label{S2hat}
  \hat{S}_2:= - \frac{1}{2}n + \frac{1}{2M} \sum_{j=1}^M \alpha_j
\end{equation}
as an approximation for $S_2$. The corresponding error estimator is
\begin{equation} \label{E2}
 E_2= \exp(- \frac{1}{n} \hat{S}_2).
 \end{equation}
  For the approximation of $S_3$ we replace
step 3 in the algorithm above by \\[3mm]
\hspace*{1cm} 3.~ $y_j:={\mathcal E}_{E^l}z_j, \quad \hat{y}_j:=
{\mathcal E}_{E^l}y_j, \quad \alpha_j:=\frac{3}{2}y_j^Ty_j- \frac{1}{3}
y_j^T\hat{y}_j  $ \\[3mm]
and we use 
\begin{equation} \label{S3hat}
 \hat{S}_3:= - \frac{7}{6}n + \frac{1}{M} \sum_{j=1}^M \alpha_j
\end{equation}
as an estimate for $S_3$. The corresponding error estimator is 
\begin{equation} \label{E3}
 E_3=\exp(- \frac{1}{n} \hat{S}_3)~.
 \end{equation}
 Clearly, this technique can be extended to
the partial sums $S_m$ with $m>3$. However, in our applications we only
use $\hat{S}_2$ and $\hat{S}_3$ for error estimation. It turns out that,
at least in our experiments,
the two leading terms in the expansion (\ref{expansion}) are sufficient
for a reasonable error estimation.
Note that due to the truncation of the Taylor expansion,
the estimators $E_2$ and $E_3$ for $d({\mathcal E}_{E^l})$ are
biased.
\\\\
It is shown in
\cite{Bai2} that based  on
 the so-called Hoeffding inequality (cf. \cite{Pollard}) probabilistic bounds
for $| \frac{1}{M} \sum_{i=1}^M z_i^THz_i - \tr(H)|$ can be derived,
 where $z_1, z_2, \ldots
,z_M$ are independent random variables as in Proposition~\ref{Carlo}. 
In this paper we do not use these bounds. Based on numerical experiments
we take a fixed small value for the parameter $M$ in the Monte Carlo algorithm
above (in the experiments in \S5: $M=6$).
\begin{remark}
{\rm
In the setting of this paper Proposition~\ref{Carlo} is applied
with $H=p({\mathcal E}_{E^l})$, where $p$ is a known polynomial
of degree 2 or 3. In the Monte Carlo technique for approximating
$\det(A)=\exp(\tr(\ln (A)))$ from \cite{Bai2}, Proposition~\ref{Carlo} is applied with 
$H=\ln(A)$. The quantity $z^T \ln(A) z$, which can be considered
as a Riemann-Stieltjes integral, is approximated using 
suitable quadrature rules.
In \cite{Bai2} this quadrature is based on a Gauss-Christoffel
technique where the unknown nodes and weights in the quadrature
rule are determined using the Lanczos method. For a detailed
explanation of this method we refer to \cite{Bai2}.
\\
 A further alternative that could be considered for error estimation
is the use of this method from \cite{Bai2}.  
 In the setting here, this method could be used to
 compute a
(rough) approximation of $\det(G_{E^l}AG_{E^l}^T)^{1/n}$. We did not investigate
this possibility. The results in \cite{Bai,Bai2} give an indication that
this alternative is probably much more expensive than the method presented in
this section. }
\qquad $\Box$ \end{remark}
\section{Numerical experiments}
\label{Experiments}
In this section we present some results of numerical experiments with the methods
introduced in \S\ref{SPAI} and \S\ref{ErrorEstimation}. All experiments are done
using a MATLAB implementation. We use the MATLAB notation $nnz(B)$ for the number of
nonzero entries in a matrix $B$. \\\\
{\bf Experiment 1} (discrete 2D Laplacian). We consider the standard 5--point
discrete Laplacian on a uniform square grid with $m$ mesh points in both directions,
i.e. $n=m^2$. For this symmetric positive definite matrix the eigenvalues
are known:
\[
 \lambda_{\nu \mu}=4(m+1)^2 \left( \sin^2(\frac{\nu \pi}{2(m+1)}) +
        \sin^2(\frac{\mu \pi}{2(m+1)}) \right), \quad 1 \leq \nu,\mu \leq m ~.
\]
For the choice of the sparsity pattern $E^l$ we use a simple approach 
based on the nonzero structure of (powers of) the matrix $A$:
\begin{equation}
\label{LaplaceE}
E^l(k):=\{ (i,j)~|~ i \geq j ~~{\rm and}~~ (A^k)_{ij} \neq 0 \}~, \quad k=1,2, \ldots.
\end{equation}
We first describe some features of the methods for the case $m=30$, $k=2$ and 
after that we will vary $m$ and $k$. Let $A$ denote the discrete Laplacian for
the case $m=30$ and $L_A$ its lower triangular part. We then  have
$nnz(L_A)=2640$. For the sparse approximate inverse
we obtain $nnz(G_{E^l(2)})=6002$. The systems 
$P_iAP_i^T \hat{g}_i=(0,0,\ldots,0,1)^T$ that have to be solved to determine 
$G_{E^l(2)}$ (cf. (\ref{Practice})) have dimensions between 1 and 7; the mean
of these dimensions is 6.7. As an approximation of $d(A)=3.1379~10^3$ we obtain
\[
 d(G_{E^l(2)})^{-2}=d(\hat{G}_{E^l(2)})^{-1}=\prod_{i=1}^n 
 (\hat{G}_{E^l(2)})_{ii}^{-\frac{1}{n}}= 3.2526~10^3~.
\]
Hence $d(A)/d(G_{E^l(2)})^{-2}=0.965$. For the computation of this approximation
along the lines as described in Remark~\ref{properties}, item 3, we have to compute
the Cholesky factorizations  $P_iAP_i^T=L_iL_i^T,~ i=1,2, \ldots,n$. For this
approximately $41~10^3$ flops are needed (in the MATLAB implementation). If we compare
this with the costs of one matrix--vector multiplication 
$A\ast x$ (8760 flops), denoted by MATVEC,
it follows that for computing this approximation of $d(A)$, with
an error of 3.5 percent, we need arithmetic work comparable to only
5 MATVEC.

 We will see that the arithmetic costs for
 error estimation are significantly higher. 
 We first consider the methods of \S\ref{method1}. The arithmetic costs 
 are measured in terms of MATVEC. For the computation
 of $\alpha$ as indicated in Lemma~\ref{Mmatrix} with $\eta=0.2$, using the
  CG method with starting vector
  ${\bf 1}=(1,1,\ldots,1)^T$ we need 8  iterations. In each CG iteration we have to compute
  a matrix--vector multiplication $G_{E^l}AG_{E^l}^Tx$, which costs
  approximately 3.7 MATVEC. We obtain $\alpha_{\rm CG}=0.0155$. For the
  method based on the Lanczos method for approximating 
  $\lambda_{\rm min}(G_{E^l}AG_{E^l}^T)$ we use the heuristic
  stopping criterion
  \begin{equation} \label{stop}
   |\lambda_1^{(j)}-\lambda_1^{(j-1)}| < 0.01 |\lambda_1^{(j)}| ~.
   \end{equation}
  We then need 7 Lanczos iterations, resulting in $\alpha_{\rm Lanczos}=
  0.0254$.       A direct computation results in
   $\lambda_{\rm min}(G_{E^l}AG_{E^l}^T)=0.025347$. \\
   For the computation of $\mu=\|G_{E^l}AG_{E^l}^T\|_F^2$ we first
   computed the lower triangular part of $X=G_{E^l}AG_{E^l}^T$ and
   then computed $\|X\|_F$ (making use of symmetry). The total costs
   of this are approximately 18 MATVEC.
    Application of Lemma~\ref{Golubhere}, with $\alpha_{\rm CG} $ and
    $\alpha_{\rm Lanczos}$ yields the two intervals
  \[
   [0.880,1] \quad \mbox{and} \quad  [0.895,1]~,
  \]
  which both contain the exact error 0.965. In both cases, the total
  costs for error estimation are 40--45 MATVEC, which is approximately
  10 times more than the costs for computing the approximation
  $d(G_{E^l(2)})^{-2}$. 
    
   We now consider the method of \S\ref{method2}. We use the estimators
   $E_2$ and $E_3$ from (\ref{E2}), (\ref{E3}) with $M=6$. The
   results are $E_2=0.980$, $E_3=0.973$. Note that the order
   of magnitude of the exact error ($3.5$ percent) is approximated well by
   both $E_2$ ($2.0$ percent) and $E_3$ ($2.7$ percent). 
   In step 3 in the Monte Carlo algorithm for computing
   $\hat{S}_2$ we need one matrix--vector
   multiplication $G_{E^l}AG_{E^l}^Tx$
   (costs  3.7 MATVEC). The total arithmetic costs for
   $E_2$ are approximately 20 MATVEC. For $\hat{S}_3$ we need two matrix--vector
   multiplications with ${\mathcal E}_{E^l}$ in the third step of
   the Monte Carlo algorithm. The total costs for $E_3$ are
   approximately 40 MATVEC. 
   
   In Table~\ref{laplace1} we give results for the discrete 2D
   Laplacian with $m=30$ ($n=900$), $m=100$ ($n=10000$) and $m=200$
   ($n=40000$). We use the sparsity pattern $E^l(2)$. In the third column
   of this table we give the computed approximation of $d(A)$ and the
   corresponding relative error. In the fourth column we give the total
   arithmetic costs for the Cholesky factorization of the matrices
   $P_iAP_i^T$, $i=1,2,\ldots,n$ (cf. Remark~\ref{properties}, item 3).
   In the columns 5--8 we give the results and corresponding arithmetic
   costs for the error estimators discussed in \S\ref{ErrorEstimation}.
   The fifth column corresponds to the method discussed in \S\ref{method1}
   with $\alpha$ determined using the CG method applied to
   $G_{E^l}AG_{E^l}^T={\bf 1}$ with starting vector $\bf 1$. 
   In the stopping criterion we take $\eta=0.2$ (cf. Lemma~\ref{Mmatrix}).
   The computed $\alpha=\alpha_{\rm CG}$ is used as input for
   the lower bound in (\ref{errorbounds}). The resulting bound for the
   relative error and the arithmetic costs  for computing
   this error bound are shown in column 5. In column 6 one finds the
   computed error bounds if $\alpha$ is determined using the Lanczos
   method  with stopping criterion (\ref{stop}). In the last two columns
   the results for the Monte Carlo estimators $E_2$ (\ref{E2}) and
   $E_3$ (\ref{E3}) are given.  
\begin{table} 
\caption{Results for 2D discrete Laplacian with $E^l=E^l(2)$} 
\begin{center} \footnotesize
\begin{tabular}{|c|c|c|c|c|c|c|c|} \hline  
 $n$ & $d(A)$ & $d(G_{E^l(2)})^{-2}$ & costs for & Thm.~\ref{Golubhere}, &
  Thm.~\ref{Golubhere}, & MC & MC  \\ 
    &        &  (error)&   $d(G_{E^l(2)})^{-2}$  & 
 $\alpha_{\rm CG}$ &  $\alpha_{\rm Lanczos}$ & 
  $E_2$ & $E_3$  \\ \hline 
\lower.3ex\hbox{900} & \lower.3ex\hbox{3.138 $10^3$} & 
\lower.3ex\hbox{3.253 $10^3$} &  
\lower.3ex\hbox{5 MV} & \lower.3ex\hbox{$\leq 12$\%} & 
\lower.3ex\hbox{$\leq 11$\% } &
 \lower.3ex\hbox{ 2.0\%} &    \lower.3ex\hbox{2.7\%}  \\
 &  &  \lower.3ex\hbox{(3.5\%)} & & \lower.3ex\hbox{(45 MV)}  & 
  \lower.3ex\hbox{(45 MV)} & \lower.3ex\hbox{(20 MV)} & 
  \lower.3ex\hbox{(40 MV)} \\ \hline
\lower.3ex\hbox{10000} & \lower.3ex\hbox{3.292 $10^4$} & 
\lower.3ex\hbox{3.434 $10^4$} &  
\lower.3ex\hbox{ 5 MV} & \lower.3ex\hbox{$\leq 21$\%} & 
\lower.3ex\hbox{$\leq 19$\% } &
 \lower.3ex\hbox{2.2\%} & \lower.3ex\hbox{2.6\%}  \\
  &  &  \lower.3ex\hbox{(4.1\%)} & & \lower.3ex\hbox{(140 MV)}  & 
  \lower.3ex\hbox{(102 MV)} & \lower.3ex\hbox{(24 MV)} & 
  \lower.3ex\hbox{(48 MV)} \\ \hline
\lower.3ex\hbox{40000} & \lower.3ex\hbox{1.300 $10^5$} & 
\lower.3ex\hbox{1.359 $10^5$} &  
\lower.3ex\hbox{ 5 MV} & \lower.3ex\hbox{$\leq 26$\%} & 
\lower.3ex\hbox{$\leq  24$\% } &
 \lower.3ex\hbox{2.2\%} & \lower.3ex\hbox{2.7\%}  \\
  &  &  \lower.3ex\hbox{(4.3\%)} & & \lower.3ex\hbox{(276 MV)}  & 
  \lower.3ex\hbox{(159 MV)} & \lower.3ex\hbox{(24 MV)} & 
  \lower.3ex\hbox{(48 MV)} \\ \hline
\end{tabular}
\end{center}  
\label{laplace1}
\end{table}  
In Table~\ref{laplace2} we show the results and corresponding
arithmetic costs for the method with sparsity pattern $E^l=E^l(4)$.
\begin{table} 
\caption{Results for 2D discrete Laplacian with $E^l=E^l(4)$} 
\begin{center} \footnotesize
\begin{tabular}{|c|c|c|c|c|c|c|c|} \hline  
 $n$ & $d(A)$ & $d(G_{E^l(4)})^{-2}$ & costs for & Thm.~\ref{Golubhere}, &
  Thm.~\ref{Golubhere}, & MC & MC  \\ 
    &        &  (error)&   $d(G_{E^l(4)})^{-2}$  & 
 $\alpha_{\rm CG}$ &  $\alpha_{\rm Lanczos}$ & 
  $E_2$ & $E_3$  \\ \hline 
\lower.3ex\hbox{900} & \lower.3ex\hbox{3.138 $10^3$} & 
\lower.3ex\hbox{3.177 $10^3$} &  
\lower.3ex\hbox{41 MV} & \lower.3ex\hbox{$\leq 3.5$\%} & 
\lower.3ex\hbox{$\leq 3.0$\% } &
 \lower.3ex\hbox{ 0.65\%} &    \lower.3ex\hbox{1.1\%}  \\
 &  &  \lower.3ex\hbox{(1.2\%)} & & \lower.3ex\hbox{(137 MV)}  & 
  \lower.3ex\hbox{(146 MV)} & \lower.3ex\hbox{(54 MV)} & 
  \lower.3ex\hbox{(108 MV)} \\ \hline
\lower.3ex\hbox{10000} & \lower.3ex\hbox{3.292 $10^4$} & 
\lower.3ex\hbox{3.347 $10^4$} &  
\lower.3ex\hbox{ 45 MV} & \lower.3ex\hbox{$\leq 7.7$\%} & 
\lower.3ex\hbox{$\leq 7.0$\% } &
 \lower.3ex\hbox{0.91\%} & \lower.3ex\hbox{1.1\%}  \\
  &  &  \lower.3ex\hbox{(1.6\%)} & & \lower.3ex\hbox{(263 MV)}  & 
  \lower.3ex\hbox{(226 MV)} & \lower.3ex\hbox{(55 MV)} & 
  \lower.3ex\hbox{(110 MV)} \\ \hline
\lower.3ex\hbox{40000} & \lower.3ex\hbox{1.300 $10^5$} & 
\lower.3ex\hbox{1.323 $10^5$} &  
\lower.3ex\hbox{46 MV} & \lower.3ex\hbox{$\leq 10$\%} & 
\lower.3ex\hbox{$\leq  9.3 $\% } &
 \lower.3ex\hbox{0.93\%} & \lower.3ex\hbox{1.1\%}  \\
  &  &  \lower.3ex\hbox{(1.7\%)} & & \lower.3ex\hbox{(487 MV)}  & 
  \lower.3ex\hbox{(348 MV)} & \lower.3ex\hbox{(56 MV)} & 
  \lower.3ex\hbox{(112 MV)} \\ \hline
\end{tabular}
\end{center}  
\label{laplace2}
\end{table}
\ \\  
Concerning the numerical results we note the following. From the third and
fourth column in Table~\ref{laplace1} we see that using this
method  we can obtain
an approximation of $d(A)$ with relative error only a few
percent and arithmetic costs only a few MATVEC. Moreover, this efficiency
hardly depends on the dimension $n$. Comparison of the third and fourth
columns of the Tables~\ref{laplace1} and \ref{laplace2} shows that
the approximation  significantly improves if we enlarge the
pattern from $E^l(2)$ to $E^l(4)$. The corresponding arithmetic costs 
increase by a factor of about 9. This is caused by the fact that the
mean of the dimensions of the systems $P_iAP_i^T$, $i=1,2,\ldots,n$,
increases from approximately 7 ($E^l(2)$) to approximately 20.
For $n=10000$ we have $nnz(L_A)=29800,~ nnz(G_{E^l(2)})=69002,
~nnz(G_{E^l(4)})=204030$. For
the other $n$ values we have similar ratios between the number of
 nonzeros in the matrices $L_A$ and $G_{E^l}$. Note that the matrix
 $G_{E^l}$ has to be stored for the error estimation but not for
 the computation of the approximation $d(G_{E^l})^{-2}$. The error
 bounds in the fifth and sixth column in the
 Tables~\ref{laplace1} and \ref{laplace2}
 are rather conservative and expensive. Furthermore there is some
 deterioration in the quality and a quite strong increase in the
 costs if the dimension $n$ grows. The strong increase in the
 costs is mainly due to the fact that the  CG and Lanczos method
 both need significantly more iterations if $n$ increases. This is
 a well-known phenomenom (the matrix $G_{E^l}AG_{E^l}^T$ has 
 a condition number that is proportional to $n$). Also note that
 the costs for these error estimators are (very) high compared to the
 costs of the computation of $d(G_{E^l})^{-2}$. The results
 in the last two columns indicate that the Monte Carlo
 error estimators, although less
 reliable, are more favourable.
 
\begin{figure}[ht] 
\caption{ Eigenvalues of the matrix $G_{E^l}AG_{E^l}^T$ in Experiment 1} 
\begin{center}
\epsfig{file=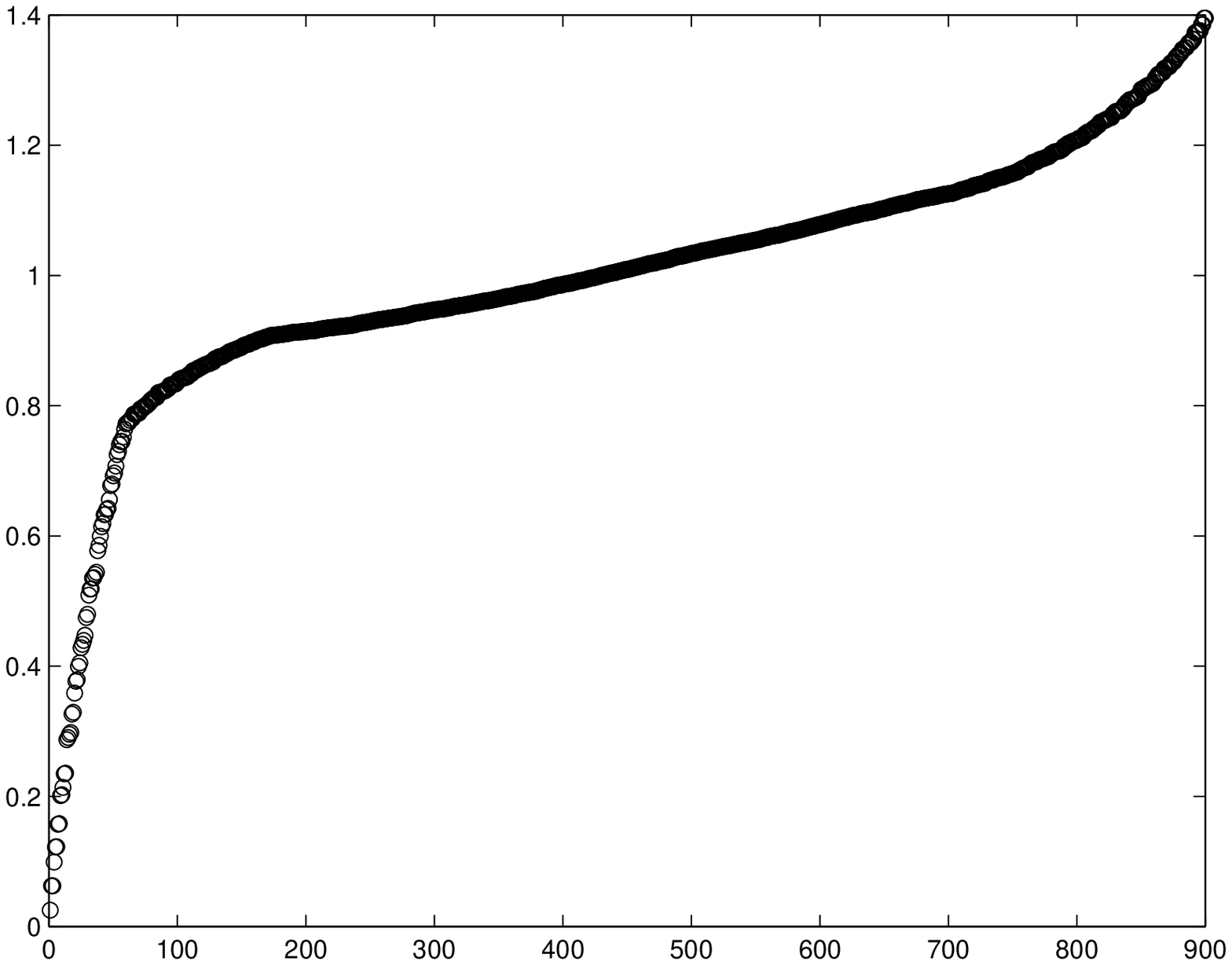,height=6cm,width=8cm}
\end{center}
\label{eigLaplace} 
\end{figure}

\begin{figure}[ht] 
\caption{ Eigenvalues of the matrices $A$ and $G_{E^l}AG_{E^l}^T$ in Experiment 2} 
\begin{center}
 \epsfig{file=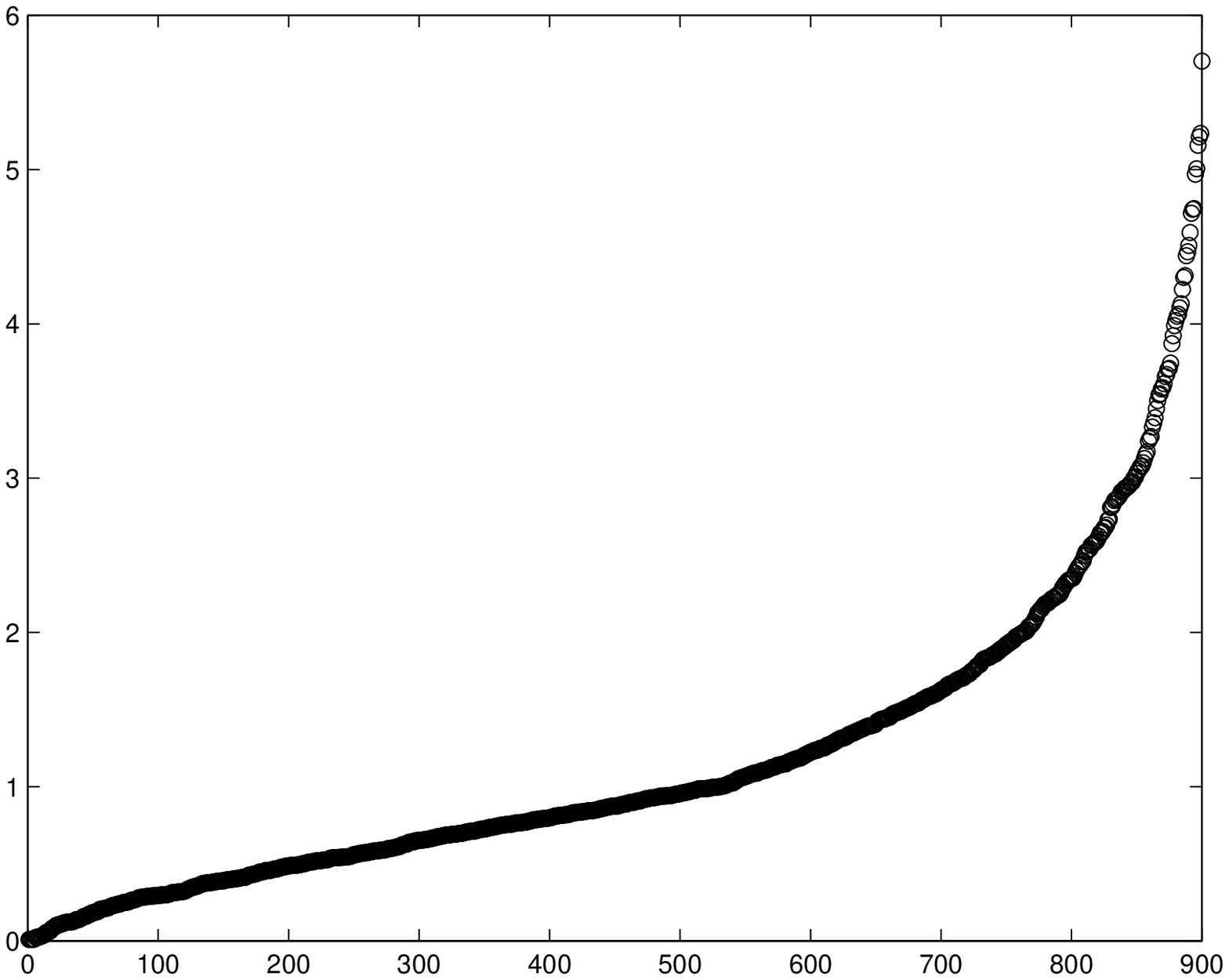,width=0.46\textwidth} \hspace*{0.4cm}
 \epsfig{file=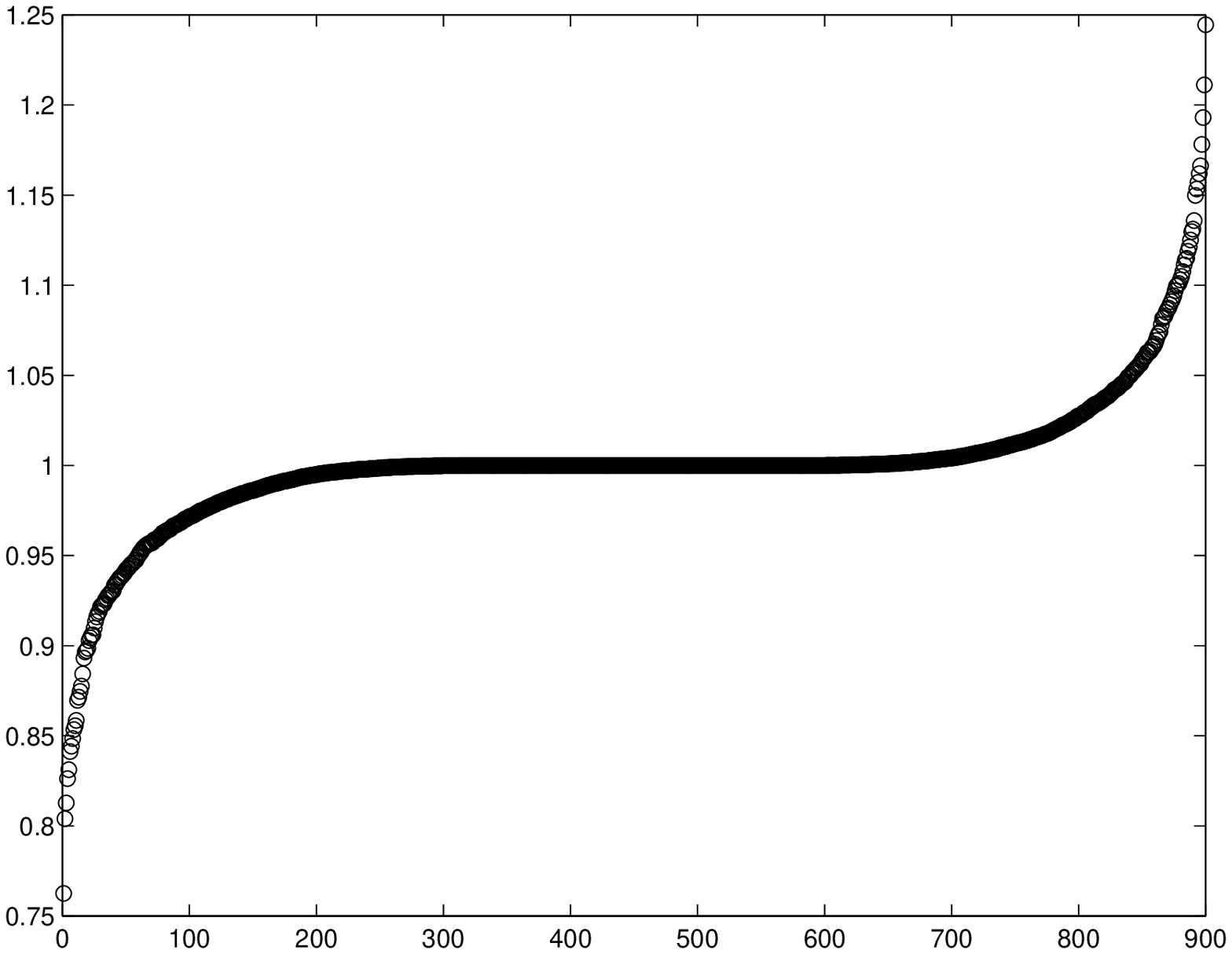,width=0.46\textwidth}
\end{center}
\label{eigRandom} 
\end{figure}

\begin{figure}[ht] 
\caption{ Eigenvalues of the matrices $A$ and $G_{E^l}AG_{E^l}^T$ in Experiment 3} 
\begin{center}
 \epsfig{file=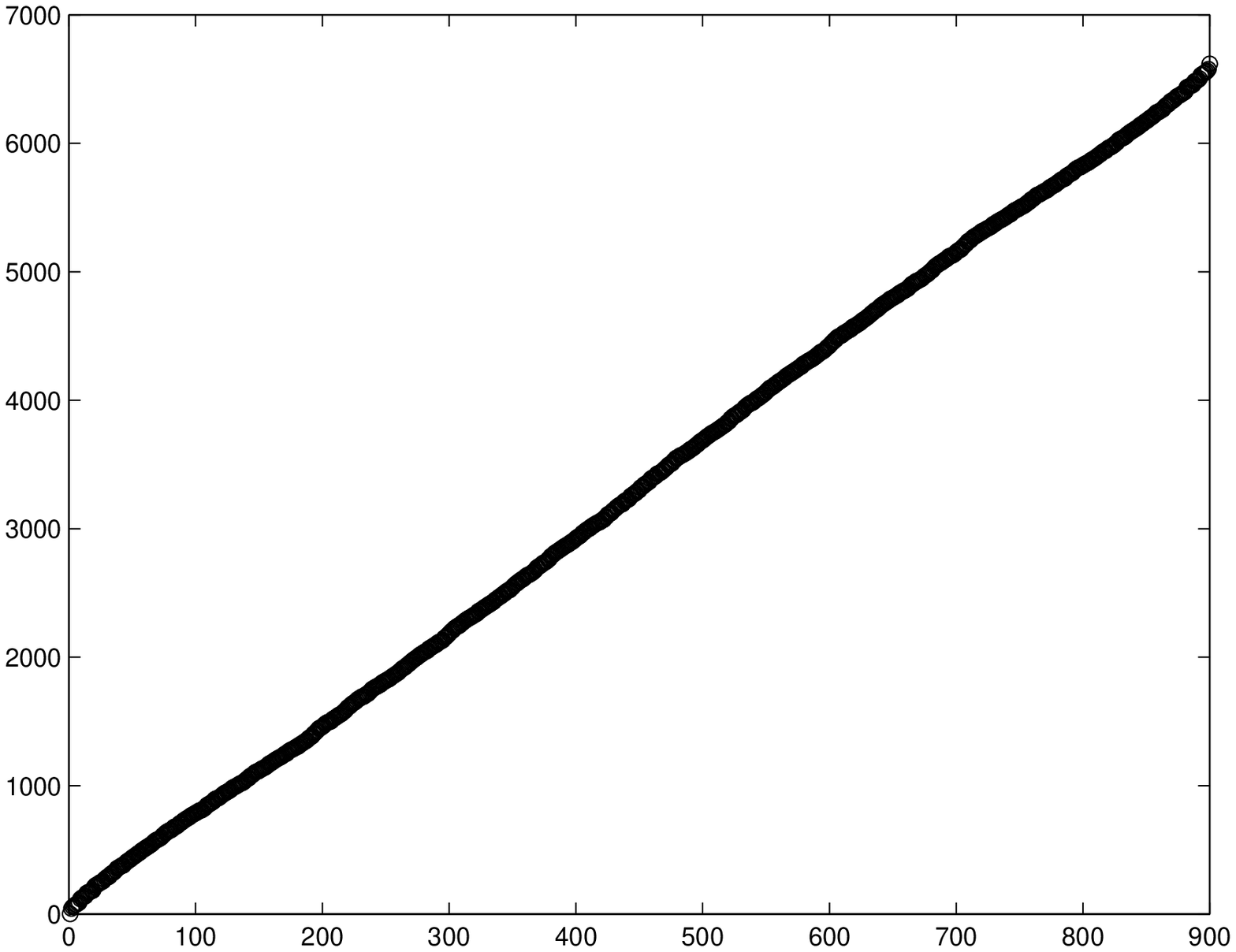,width=0.46\textwidth} \hspace*{0.4cm}
 \epsfig{file=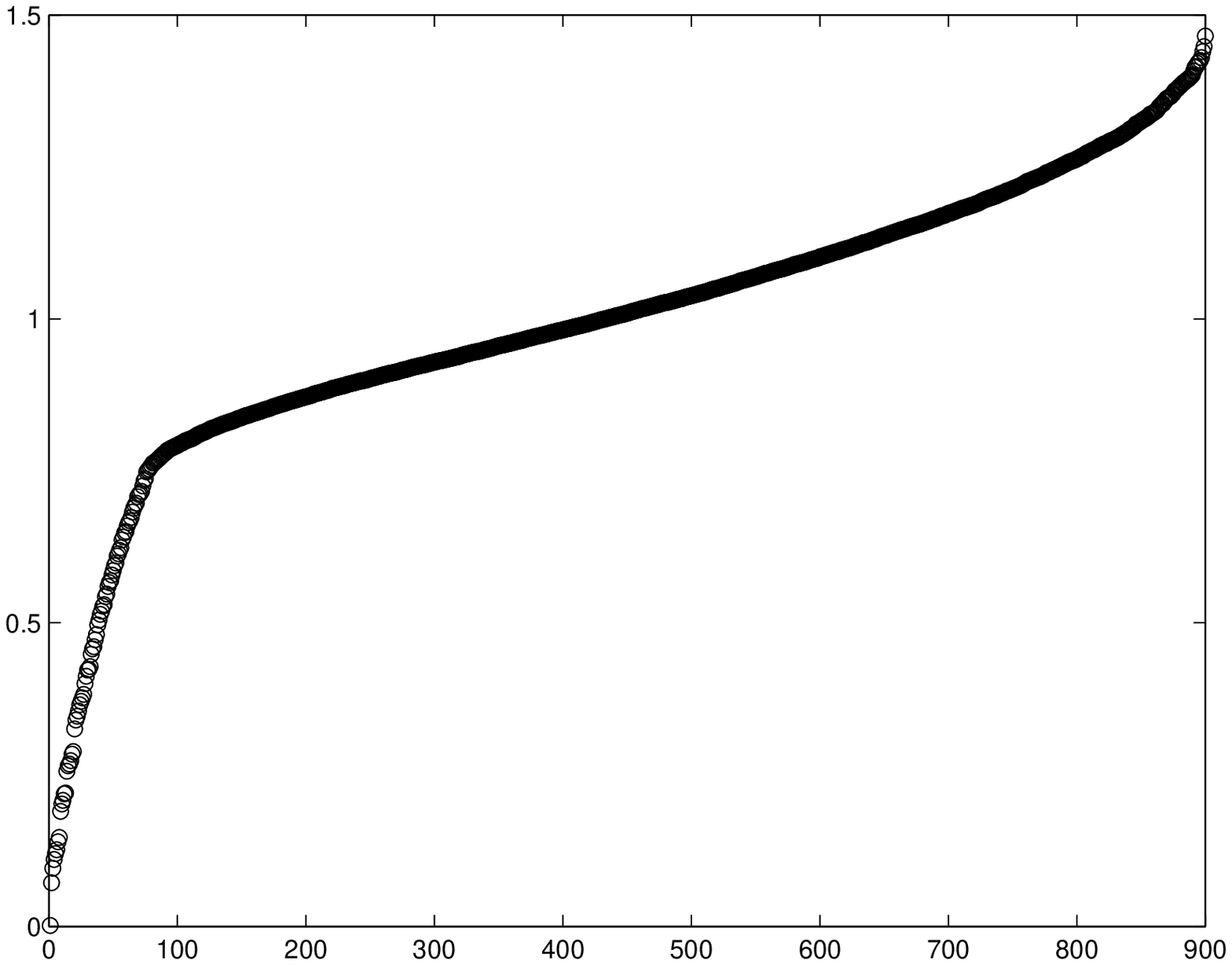,width=0.46\textwidth}
\end{center}
\label{FigQCD} 
\end{figure}

 In Figure~\ref{eigLaplace} we show the eigenvalues of the
 matrix $G_{E^l}AG_{E^l}^T$ for the case $n=900$, $E^l=E^l(2)$
 (computed with the MATLAB function {\sc eig}). The eigenvalues
 are in the interval $[0.025, 1.4]$.
  The mean
 of these eigenvalues is 1 ( $\tr(G_{E^l}AG_{E^l}^T)=1$). One can
 see that relatively many eigenvalues are close to 1 and only a few
 eigenvalues are close to zero.   
\ \\\\
{\bf Experiment 2} (MATLAB random sparse matrix). The sparsity
structure of the matrices considered
in Experiment 1 is  very regular. In this experiment we consider
matrices with a pattern of  nonzero entries that  is very
irregular. We used the MATLAB generator ({\sc sprand}$(n,n,2/n)$) to
generate a  matrix $B$ of order $n$ with approximately $2n$ nonzero
entries. These are uniformly distributed random entries in $(0,1)$.
The matrix $B^TB$ is then
sparse symmetric positive semidefinite. In the generic case this matrix
has many eigenvalues zero. To obtain a positive definite
matrix we generated a random vector $d$ with all
entries chosen from a uniform distribution
on the interval  $(0,1)$ ($d:=${\sc rand}$(n,1)$). As a testmatrix we
 used $A:=B^TB+{\rm diag}(d)$. We  performed
  numerical experiments similar to those in
Experiment 1 above. We only consider the case with sparsity pattern
$E^l=E^l(2)$. The error estimator based on the CG method is not
applicable because the sign condition in Lemma~\ref{Mmatrix} is not
fulfilled.  For the case $n=900$ the eigenvalues
of $A$ and of $G_{E^l}AG_{E^l}^T$ are shown in Figure~\ref{eigRandom}. For $A$ the smallest and largest
eigenvalues are $0.0099$ and $5.70$, respectively. The picture on the
right in Figure~\ref{eigRandom} shows that for this matrix $A$
 sparse approximate inverse
preconditioning results in a very well--conditioned matrix. Related to this, one can see in 
Table~\ref{Tablerand} that for this random matrix $A$ the approximation
of $d(A)$ based on the sparse approximate inverse is much better
than for the discrete Laplacian in Experiment 1. 
For $n=900, 10000, 40000$ we obtain
$nnz(L_A)= 2730, 29789, 120216$ and $nnz(G_{E^l})= 7477, 82290,
335139$, respectively. For $n=900, 10000, 40000$ the mean of the
dimensions of the systems $P_iAP_i^T$, $i=1,2,\ldots,n$ is 
$10.6, 10.8, 11.0$, respectively. In all three cases  the costs for a
matrix--vector multiplication $G_{E^l}AG_{E^l}x$ are approximately
4.3 MV. Furthermore, in all three cases the matrix  $G_{E^l}AG_{E^l}^T$
is well--conditioned and the number of Lanczos iterations
needed to satisfy the stopping criterion (\ref{stop})
hardly depends on $n$. Due to this, for increasing $n$,
the growth in  the costs for the error estimator based on Theorem~\ref{Golubhere}
(column 5) is much slower than in Experiment 1.
As in the  Tables~\ref{laplace1} and \ref{laplace2},
 in Table~\ref{Tablerand}
the error quantities in the columns
 3, 5,6,7 are bounds or estimates for the relative
 error $1-d(G_{E^l}AG_{E^l}^T)$.\\
 For $n=10000, 40000$ the values of $d(A)$ are not given (column 2).
 This has to do with the fact that for these matrices with very irregular
 sparsity patterns the Cholesky factorization $A=LL^T$
 suffers from much
 more fill-in than for the matrices in the Experiments 1 and 3. For
  the matrix $A$ in this experiment with
 $n=900$ we have $nnz(L_A)=2730$ and $nnz(L)=72766$. For $n=10000$ we
 run into storage problems if we try to compute the 
 Cholesky factorization using the MATLAB function {\sc chol}. \\\\  
\begin{table} 
\caption{Results for MATLAB random sparse matrices with $E^l=E^l(2)$} 
\begin{center} \footnotesize
\begin{tabular}{|c|c|c|c|c|c|c|} \hline  
 $n$ & $d(A)$ & $d(G_{E^l})^{-2}$ & costs for & Thm.~\ref{Golubhere},  
 & MC & MC  \\ 
    &        &  (error)&   $d(G_{E^l})^{-2}$  & 
   $\alpha_{\rm Lanczos}$ & 
  $E_2$ & $E_3$  \\ \hline 
\lower.3ex\hbox{900} & \lower.3ex\hbox{0.82453} & 
\lower.3ex\hbox{0.82521} &  
\lower.3ex\hbox{23 MV}  & 
\lower.3ex\hbox{$\leq 9.8~10^{-4}$ } &
 \lower.3ex\hbox{1.4 $10^{-3}$} &    \lower.3ex\hbox{1.0 $1.0^{-3}$}  \\
 &  &  \lower.3ex\hbox{(8.3 $10^{-4}$)} &  & 
  \lower.3ex\hbox{(110 MV)} & \lower.3ex\hbox{(26 MV)} & 
  \lower.3ex\hbox{(52 MV)} \\ \hline
\lower.3ex\hbox{10000} & \lower.3ex\hbox{--} & 
\lower.3ex\hbox{0.81053 } &  
\lower.3ex\hbox{18  MV}  & 
\lower.3ex\hbox{$\leq 1.1~10^{-3}$ } &
 \lower.3ex\hbox{8.4 $10^{-4}$} & \lower.3ex\hbox{7.4 $10^{-4}$}  \\
  &  &  \lower.3ex\hbox{(--)} &   & 
  \lower.3ex\hbox{(139 MV)} & \lower.3ex\hbox{(26 MV)} & 
  \lower.3ex\hbox{(52 MV)} \\ \hline
\lower.3ex\hbox{40000} & \lower.3ex\hbox{--} & 
\lower.3ex\hbox{0.82033} &  
\lower.3ex\hbox{18 MV} & 
\lower.3ex\hbox{$\leq  1.0~10^{-3}$} &
 \lower.3ex\hbox{6.2 $10^{-4}$} & \lower.3ex\hbox{8.3 $10^{-4}$}  \\
  &  &  \lower.3ex\hbox{(--)} &   & 
  \lower.3ex\hbox{(146 MV)} & \lower.3ex\hbox{(26 MV)} & 
  \lower.3ex\hbox{(52 MV)} \\ \hline
\end{tabular}
\end{center}  
\label{Tablerand}
\end{table}  
{\bf Experiment 3} (QCD type matrix). In this experiment we consider
a complex Hermitean positive definite matrix with sparsity structure
as in Experiment 1. This matrix is 
motivated by applications from the QCD field.
 In QCD simulations the determinant
of the  so-called Wilson fermion
matrix is of interest. These matrices and some of their
properties are discussed in \cite{Forcrand,Frommer}.  The nonzero entries in 
a Wilson fermion  matrix are induced
by a nearest neighbour coupling in a regular 4-dimensional grid. These
couplings consist of $12 \times 12$ complex matrices $M_{xy}$, which have
a tensor product structure $M_{xy}=P_{xy} \otimes U_{xy}$, where $P_{xy} \in
 \Bbb{R}^{4 \times 4}$
is a  projector, $U_{xy} \in \Bbb{C}^{3 \times 3}$ is from $SU_3$ and $x$ and $y$
denote nearest neighbours in the grid.
These coupling matrices $M_{xy}$ strongly fluctuate as a function of $x$ and $y$.
Here we consider a (toy) problem with a matrix
which  has some similarities with these Wilson fermion matrices. We start
with a 2-dimensional regular grid as in Experiment 1 ($n$
grid points). For the
couplings with nearest neighbours we use complex numbers with length 1.
These numbers are chosen as follows. The couplings with south and
west neighbours at a grid point $x$
are $\exp(2i \pi \alpha_S(x))$ and $\exp(2i \pi \alpha_W(x))$,
respectively, where $\alpha_S(x)$ and $\alpha_W(x)$ are chosen
from a uniform distribution on the interval $(0,1)$.
The couplings with the north and east neighbours are taken such that the matrix
is hermitean.  To make the comparison with  Experiment 1 easier
the matrix is scaled by the factor $n$, i.e. the
couplings with nearest neighbours have length $n$.
For the diagonal we take $\gamma I$, where
$\gamma$ is chosen such that the smallest eigenvalue of the
resulting matrix is approximately 1 (this can be realized by using the MATLAB 
function {\sc eigs} for estimating the smallest eigenvalue). We performed
numerical experiments as in Experiment 1 with $E^l=E^l(2)$.
The number of nonzero entries in $L_A$ and $G_{E^l}$ are the same
as in Experiment 1. For $n=900$ the eigenvalues of the matrices
$A$ and $G_{E^l}AG_{E^l}^T$ are shown in Figure~\ref{FigQCD}. These
spectra are in the
 intervals $[1,~6.6~10^3]$ and $[1.7~10^{-3},~1.5]$, respectively.
 \\  
The results of numerical experiments are presented in Table~\ref{TableQCD}.
Note that the error estimator from \S\ref{method1}
in which the CG method is used for computing $\alpha$
can not be used for this matrix (assumptions in Lemma~\ref{Mmatrix}
are not satisfied). We did not consider the case $n=40000$ here
because then the application of the {\sc eig} function for computing
the smallest eigenvalue led to memory problems. \\
\begin{table}[ht]
\caption{Results for QCD type matrix with $E^l=E^l(2)$} 
\begin{center} \footnotesize
\begin{tabular}{|c|c|c|c|c|c|c|} \hline  
 $n$ & $d(A)$ & $d(G_{E^l})^{-2}$ & costs for & Thm.~\ref{Golubhere},  
 & MC & MC  \\ 
    &        &  (error)&   $d(G_{E^l})^{-2}$  & 
   $\alpha_{\rm Lanczos}$ & 
  $E_2$ & $E_3$  \\ \hline 
\lower.3ex\hbox{900} & \lower.3ex\hbox{2.500 $10^3$} & 
\lower.3ex\hbox{2.620 $10^3$} &  
\lower.3ex\hbox{5 MV}  & 
\lower.3ex\hbox{$\leq 24$\%} &
 \lower.3ex\hbox{2.6\%} &    \lower.3ex\hbox{3.3\%}  \\
 &  &  \lower.3ex\hbox{(4.6\%)} &  & 
  \lower.3ex\hbox{(79 MV)} & \lower.3ex\hbox{(23 MV)} & 
  \lower.3ex\hbox{(46 MV)} \\ \hline
\lower.3ex\hbox{10000} & \lower.3ex\hbox{2.739 $10^4$} & 
\lower.3ex\hbox{2.842 $10^4$ } &  
\lower.3ex\hbox{5  MV}  & 
\lower.3ex\hbox{$\leq 31$\% } &
 \lower.3ex\hbox{2.4\%} & \lower.3ex\hbox{2.7\%}  \\
  &  &  \lower.3ex\hbox{(3.6\%)} &   & 
  \lower.3ex\hbox{(133 MV)} & \lower.3ex\hbox{(23 MV)} & 
  \lower.3ex\hbox{(46 MV)} \\ \hline
\lower.3ex\hbox{22500} & \lower.3ex\hbox{6.173 $10^4$} & 
\lower.3ex\hbox{6.391 $10^4$} &  
\lower.3ex\hbox{5 MV} & 
\lower.3ex\hbox{$\leq  32$\%} &
 \lower.3ex\hbox{2.3\%} & \lower.3ex\hbox{2.8\%}  \\
  &  &  \lower.3ex\hbox{(3.4\%)} &   & 
  \lower.3ex\hbox{(198 MV)} & \lower.3ex\hbox{(23 MV)} & 
  \lower.3ex\hbox{(46 MV)} \\ \hline
\end{tabular}
\end{center}  
\label{TableQCD}
\end{table}  
Comparison of the results in Table~\ref{TableQCD} with those in 
Table~\ref{laplace1} shows 
that when the method is applied to the QCD type of problem
instead of the discrete Laplacian 
 the performance of the method does not change very much. 
 
Finally, we note that in all measurements of the arithmetic costs we did
not take into account  the costs of determining the sparsity pattern
$E^l(k)$ and of building the matrices $P_iAP_i^T$.

\end{document}